\documentclass[aps,prc,superscriptaddress,twocolumn]{revtex4-2}

\usepackage{graphicx}
\usepackage{amsmath}

\usepackage{amssymb}
\usepackage{bm}

\usepackage[utf8]{inputenc}
\usepackage[T1]{fontenc}
\begin{document}

\title{Neutron Optical Constraints on Exotic Parity-odd Interactions between Neutrons and Nucleons}

\author{V. Kurmangaliyeva}
\author{O. Agyl-Mussapar}
\author{S. Amangeldinova}
\author{B. Massak}

\affiliation{al-Farabi Kazakh National University, Almaty, Kazakhstan}

\author{W. M. Snow}
\affiliation{Indiana University/CEEM, 2401 Milo B. Sampson Lane, Bloomington, Indiana 47408, USA}

\date{\today}

\begin{abstract}

Many theories beyond the standard model of particle physics predict the existence of weakly-coupled exotic spin-dependent interactions between standard model particles. We use experimental upper bounds on neutron-nucleus parity violation at meV and keV neutron energies in the forward scattering limit described by neutron optics to constrain exotic parity-odd interactions of the neutron with nucleons from exotic vector boson exchange for force ranges between $10^{-2}$ to $10^{-14}$ meters.

\end{abstract}

\maketitle

\section {Introduction}

Many theories beyond the standard model of particle physics predict the existence of weakly-coupled exotic spin-dependent interactions between standard model particles~\cite{Leitner64,Weinberg72,Moody84} mediated by spin-0 or spin-1 boson exchange. These interactions can arise from various physical mechanisms, such as the presence of pseudo-Goldstone bosons resulting from incompletely broken symmetries~\cite{Weinberg72}, compactifications in string theory~\cite{string1,string2,string3}, and other theoretical frameworks~\cite{paraphoton1,paraphoton2,majorons1,zboson1,zboson2,s1boson1,s1boson2,s1boson3,s1boson4,s1boson5}. 

The possible forms of such interactions are highly constrained within quantum field theory. In this case the possible parameters describing such interactions are limited to the exchange boson masses and spins and the various possible coupling constants: scalar and pseudoscalar couplings for spin-0 boson exchange, and vector and axial-vector couplings for spin-1 boson exchange. Since the strength of these interactions is quite weak, the amplitude from boson exchange can be calculated perturbatively and can be analyzed typically as a small perturbation on top of the standard model interactions.  For the large subset of experiments which employ measurements among nonrelativistic nucleons and electrons, one can write down a complete set of nonrelativistic potentials involving the spins, momenta, interaction range, and possible couplings of the particles~\cite{dobrescu_spin-dependent_2006,fadeev_revisiting_2019}. Various reviews~\cite{Adelberger_rev,Antoniadis_rev,Safronova_rev, Cong_rev} describe the broad array of experimental searches sensitive to various couplings, interaction ranges, and potentials performed to date.

Slow neutrons possess properties well suited to experimental searches for possible exotic spin-dependent interactions. Slow neutron beams can be polarized, analyzed, and detected with high efficiency. The small electric polarizability of neutrons compared to atoms and the ability of slow neutrons to penetrate macroscopic amounts of dense matter with negligible attenuation and decoherence enables one to realize sensitive searches for spin-dependent effects.  Polarized slow neutron beams have set some of the most sensitive limits on many types of weakly coupled spin-dependent interactions~\cite{Piegsa2012,Yan13,Lehnert17, Parnell:2020, Mulkey:2025wnb}. Typically these limits are the most stringent limits available for exotic neutron spin-dependent interactions with interaction ranges from 1mm down to distance scales near the size of the nucleus. 

Experimental searches for exotic spin-dependent interactions involving a polarized species share several features with experiments performed to measure parity violation from electroweak standard model interactions. In both cases the relevant experimental observable is typically chosen to be a spin-dependent asymmetry coming from the interference between parity-even interactions and the parity-odd component of the electroweak interaction proportional to $\vec{\sigma} \cdot \vec{p}$ where $\vec{\sigma}$ is a particle spin and $\vec{p}$ is a particle momentum. A very wide variety of measurements of this type have been conducted  using polarized neutrons, polarized electrons, polarized protons, and polarized photons interacting with unpolarized targets. So far all of the nonzero observations of parity-odd effects in these experiments are well described by the standard electroweak model. This agreement between theory and experiment constrains possible exotic parity-odd effects. 

The analysis of parity-odd electron-nucleon and electron-electron interactions can often be performed with high theoretical precision. By contrast our understanding of the parity-odd nucleon-nucleon interactions in the standard electroweak model suffers from both a scarcity of data with sufficient precision and from our limited understanding and computational control of QCD in the strongly interacting low energy limit. For this reason, we are not yet in a position to be able to conduct a global analysis of all of the NN weak interaction data to search for possible exotic NN parity-odd interactions. 

However we can make some progress on this problem through the analysis of experiments involving polarized low energy neutron transmission in the forward limit. In the forward limit one can analyze the data in a straightforward way using the concepts of neutron optics as constrained by the optical theorem of scattering theory. Nonrelativistic scattering theory applied to neutron interactions in matter shows that one can express the ratio $n=p^{'}/p$ of the magnitude of the in-matter momentum $p^{'}$ to the incident momentum magnitude $p$ in the forward scattering limit in terms of a neutron index of refraction $n=1+2\pi Nf(0)/p^{2}$ where $N$ is the density of scatterers in the medium and $f(\theta=0)=f(0)$ is the neutron-atom forward scattering amplitude. 

From this theory it also follows that neutron interactions in the medium which depend on the neutron spin $\vec{\sigma}$ exhibit phenomena directly analogous to optical activity and circular dichroism. Shortly after the parity-violating $V-A$ structure for the weak interaction current was established, F. C. Michel~\cite{Michel:1964} suggested that the neutron forward scattering amplitude $f(0)$ in matter could possess a parity-odd component of the form $f_{pnc}(0)=G \vec{\sigma} \cdot \vec{p}$ where $\vec{\sigma}$ is the neutron spin, $\vec{p}$ is the neutron momentum, and  $G$ is on the order of the Fermi constant $G_{F}$. In turn this interaction can produce parity-odd contribution to neutron optical activity and neutron dichroism of the form $d\phi_{pnc}/dz=-4\pi N Re(G)$ and $A_{z}=(\sigma_{+}-\sigma_{-})/(\sigma_{+}+\sigma_{-})=4\pi Im(G)/\sigma$ where $d\phi_{pnc}/dz$ is the parity-odd neutron optical rotary power and $A_{z}$ is the parity-odd neutron longitudinal asymmetry. These expressions are model-independent results that come from nonrelativistic scattering theory~\cite{Stodolsky}. 

In this paper we are interested in parity-odd interactions of the neutron which can come from exotic vector boson exchange from an interaction of the form $\mathcal{L}_{I}=\bar{\psi}(g_{V}\gamma^{\mu}+g_{A}\gamma^{\mu}\gamma_{5})\psi X_{\mu}$ where $g_{V}$ and $g_{A}$ are the vector and axial couplings of a light vector boson $X_{\mu}$ coupling to a fermion $\psi$  In the nonrelativistic limit this interaction gives rise to two interaction potentials of interest depending on both the spin and the relative momentum~\cite{Piegsa2012}: one proportional to $g_{A}^{2}\vec{\sigma}\cdot(\vec{v}\times\hat{r})$ and another proportional to $g_{V}g_{A}\vec{\sigma}\cdot\vec{v}$. In the nonrelativistic limit this Lagrangian generates a parity-odd potential of the form 

  \begin{equation}\label{eqn1}
  V(r)=\frac{g_{V}g_{A}}{2\pi}\frac{\exp{(-r/\lambda)}}{r}\vec{\sigma}\cdot\vec{v}
  \end{equation}
where $\lambda=1/m_{X}$ is the interaction range, $m_{X}$ is the mass of the vector boson, $\vec{s}=\vec{\sigma}/2$ is the spin of the polarized particle, and $r$ is the distance between the two interacting particles. This exotic interaction can also generate parity-odd neutron optical effects.

Two sensitive searches for parity-odd interaction of the neutron with nuclei in different energy regimes in the neutron optical limit have been conducted that are available for such an analysis. One experiment by the NSR collaboration sought the parity-odd neutron rotary power $d\phi \over dL$ of transversely polarized neutron transmission through $^{4}$He using slow neutrons in the meV neutron energy regime. The other experiment by the TRIPLE collaboration sought a parity-odd longitudinal asymmetry $A_{L}$ in longitudinally polarized neutron transmission through $^{232}$Th in the eV-keV neutron energy regime. Since these two nuclei have very different ratios of neutrons to protons, one can extract separate constraints on possible exotic parity-odd neutron-neutron and neutron-proton interactions.

The rest of this paper is organized as follows. Section 2 briefly reviews the relevant experimental results to date in neutron-nucleus parity violation measurements. Section 3 reviews the existing theoretical calculations for neutron parity violation in few nucleon systems and in nuclei and compares to the relevant experimental results. Section 4 derives the neutron optical parity violation predictions for the exotic parity violation interaction of interest and the corresponding constraints on the coupling constant products. Section 5 presents the constraints and compares them with other works. Section 6 evaluates the case of quantum dark forces. Section 7 states our conclusions and calls attention to ongoing work on parity violation in atoms and molecules which can improve these constraints in the future.

\section{Experimental Neutron-Nucleus Parity Violation Results}

Many experimental results for $d\phi_{pnc}/dz$ and for $A_{z}$  have been measured in a wide variety of nuclei which are much larger than the typical size of order $G$ one would expect from standard model neutron-nucleus parity violation. These effects are not exotic in origin. The large asymmetries seen in these experiments are understood to come from standard model parity violation effects as amplified near p-wave resonances in the dense forest of narrow neutron-nucleus resonances in the highly-excited $A+1$ system formed in low energy neutron-nucleus interactions, which populate states just above the neutron separation energy $S_{n}$ of the resulting compound nucleus. This amplification mechanism  on p-wave neutron-nucleus resonances comes from the mixing of s-wave and p-wave compound nucleus resonances and was well enough understood to be predicted theoretically~\cite{Sushkov1980, Sushkov1982} before it was confirmed by experiment~\cite{Forte80, Alfimenkov1983, Alfimenkov1984, Heckel1982, Heckel1984} and later measured with higher precision at the 0.7 eV p-wave resonance in $^{139}$La ~\cite{Yuan1991, Heil1999, Haseyama2002}. The Time Reversal Invariance and Parity at Low Energies (TRIPLE) collaboration~\cite{Mitchell:1999, Mitchell:2001} conducted a broad experimental survey of several heavy nuclei and measured the parity-odd helicity dependence of the total cross section for 75 different resonances at greater than $3\sigma$ statistical significance with atomic mass numbers between $93<A<130$ and near $A=230$. We exclude this data from the analyses presented in this paper.   

Two sensitive neutron optical parity-odd asymmetry measurements have been conducted which can be analyzed for the exotic interaction defined above. One experiment sought the parity-odd neutron rotary power $d\phi_{pnc}/dz$ in $n+^{4}$He using polarized slow neutrons with kinetic energies in the meV range. This few-body system does not possess any sources of enhancement of parity violation from neutron-nucleus resonances. The experimental result reported in the final analysis of this experiment, conducted at the NIST Center for Neutron Research, is $d\phi/dz =(+2.1 \pm 8.3 (stat.) +2.9/-0.2 (sys.))\times10^{-7} $ rad/m, consistent with zero~\cite{Swanson:2019}. 

The other experiment was conducted at Los Alamos National Lab by the TRIPLE collaboration. They measured the parity-odd asymmetry $A_{L}={{\sigma_{+}-\sigma_{-}}\over{{\sigma_{+}-\sigma_{-}}}}$ using longitudinally polarized neutrons created by a polarized proton-based neutron spin filter for neutron energies in the 1eV-1 keV range in several heavy nuclei. Although the main motivation for this work was their successful search for large parity-odd asymmetries on p-wave neutron-nucleus resonances as described above, they also at the same time took extensive data with neutrons energies that fell between resonances, since neutrons of all energies are emitted by the pulsed proton spallation reactions that create the neutron beam and allow for measurements as a function of neutron energy using neutron time-of-flight from the source to the detector. 

TRIPLE performed two analyses of $A_{L}$ data taken in the $^{232}$Th nucleus to search for parity violation in nonresonant neutron-nucleus scattering~\cite{Bowman:1993, Mitchell:2000} as a by-product of their measurements of parity violation in $^{232}$Th resonances~\cite{TRIPLE:1992jrn, PhysRevC.58.1236}. In their second and final paper~\cite{Mitchell:2000} they present the most sensitive experimental upper bound on neutron-nucleus parity violation in potential scattering the epithermal neutron energy range as:

\begin{equation}
P_{off}={{\sigma_{+}-\sigma_{-}}\over{\sigma_{+}-\sigma_{-}}} = (0.5 \pm 1.5) \times 10^{-6} 
\end{equation}

where the \lq\lq off-resonance\rq\rq asymmetry $P_{off}$ refers to an analysis in which they exclude the neutron energy regions where neutron-nucleus resonances are present to avoid standard model parity violation effects. We refer to their papers for the details of the analysis procedure and we will use their reported limits below in different neutron energy windows. 

Finally two new neutron-nucleus parity violation results have appeared in the last few years which we make use of in our analysis. A parity-odd asymmetry in the emission of the 2.2 MeV gamma ray relative to the incident neutron spin in $\vec{n}+p \to d+\gamma$ of $A^{np}_{\gamma}=[-3.0 \pm 1.4(stat) \pm 0.2(sys)] \times 10^{-8}$ was measured with $2\sigma$ statistical significance by the NPDGamma collaboration~\cite{NPDGamma:2018vhh}. The $n ^3$He Collaboration measured the P-odd correlation between the neutron polarization and the proton emission direction in the $\vec{n}+^{3}$He$\to ^{3}$H$+p$ reaction of $A_{PV}=[1.58 \pm 0.97 (stat) \pm 0.24 (sys)] \times 10^{-8}$~\cite{Gericke2020}. Since these two most sensitive neutron-nucleus weak interaction few body measurements to date are sensitive enough to resolve standard model parity violation, in this paper we will make use of them not to try to constrain exotic parity violation, but rather to compare to the meson exchange model theory for NN weak interactions. 

\section{Review of theory for standard model parity violation in neutron-nucleus forward scattering}

In this section we will use new experimental results on NN parity violation in few nucleon systems coupled to recent theoretical progress in the meson exchange model for standard model NN weak interactions to show that, in both cases, the experimental precision of the $n+^{4}$He and $n+^{232}$Th measurements referred to above falls comfortably below the values expected from standard model parity violation. Then we will use neutron optics theory to calculate how the exotic parity-odd interaction we consider would have appeared in these observables and use the experimental upper bounds quoted above to constrain the relevant coupling constant product $g_{V}g_{A}$ as a function of $\lambda$. 

To do this we draw upon the recent results of theoretical calculations conducted in the course of the development of the understanding of standard model neutron parity violation in few nucleon systems, where one can derive a direct relationship between the NN weak amplitudes and parity-odd observables, and updated results from older theoretical calculations of parity violation in heavy nuclei, where one must take a more phenomonological approach.  

In the standard model, the dynamical mechanism for parity odd interactions between nucleons in our energy range of interest can be formulated in terms of meson exchange, with one strong interaction parity-conserving coupling at one vertex and one parity-odd coupling at the other vertex. This approach is motivated by the fact that the range of the Standard Model parity-odd interactions among the quarks in the nucleon is orders of magnitude smaller than the nucleon size, so it can be treated as a contact interaction for energies at and below the strong interaction scale $\Lambda_{QCD}$. We will use a meson exchange approach in our work, which is sufficient for our purposes. Several theoretical works over many decades~\cite{DDH1980, Adelberger1985, Desplanques:1998ak, Schiavilla2004, Zhu2005, Schiavilla2008, Ramsey-Musolf:2006vfz, Schindler2013, Haxton:2013aca, Phillips:2014kna, Viviani2014, Gardner:2017xyl, deVries:2020iea, Gardner_prc:2022dwi, Gardner_plb:2022mxf, Muralidhara:2023zcs} have been devoted to the as-yet-unsolved problem of deriving the weak interactions between nucleons directly from the standard electroweak theory in a model-independent manner. 
 
The parity odd amplitudes involved in nuclear few body systems can be treated to an excellent approximation as perturbations in this and all low energy NN weak interaction processes. In the meson exchange picture which we employ in this paper, a hadronic weak interaction observable $(O_{PV})$ can be expressed completely in terms of six
weak meson-nucleon coupling constants $O_{PV} = a^1_\pi h^1_\pi +
a^0_\rho h^0_\rho + a^1_\rho h^1_\rho + a^2_\rho h^2_\rho + a^0_\omega
h^0_\omega + a^1_\omega h^1_{\omega} $ with coefficients $a^{\Delta I}$, where $\Delta I $ is the change in isospin are determined from theoretical calculations. 

Recent theoretical work by Gardner and Muralidhara~\cite{Muralidhara:2023zcs, Muralidhara:2024thesis} has presented a calculation of the NN weak amplitudes within the meson exchange model. This recent work used a factorization approximation method to evaluate the relevant 4-quark matrix elements evolved down from the electroweak scale into a product of separate meson-production and 
nucleon interaction terms, with the former taken from experimental data and the latter taken from lattice QCD calculations.
This work represents the first comprehensive NN weak interaction calculation from the standard model that delivers predictions for all of the NN weak couplings in the meson exchange model and is in agreement with parity-odd NN amplitudes measured in few body systems within its domain of applicability and with no adjustible parameters. It is a major advance beyond the wide ranges for the NN weak couplings long presented as best values and reasonable ranges in the Desplanques, Donoghue, and Holstein (DDH)~\cite{DDH1980} model, which was unable to predict the size of the NN weak couplings. 

The NPDGamma collaboration measurement of $A^{np}$ in $\vec{n}+p \to d+\gamma$, $A^{np}_{\gamma}=[-3.0 \pm 1.4(stat) \pm 0.2(sys)] \times 10^{-8}$~\cite{NPDGamma:2018vhh} determines the $\Delta I=1$,  $^3S_1 -{^3P_1}$ component of the weak NN interaction, dominated by the weak pion exchange coupling $h^1_\pi$ in the meson exchange picture. The $n ^3$He Collaboration measurement of $A_{PV}$ in the $\vec{n}+^{3}$He$\to ^{3}$H$+p$ reaction, $A_{PV}=[1.58 \pm 0.97 (stat) \pm 0.24 (sys)] \times 10^{-8}$~\cite{Gericke2020} constrains a linear combination of $\rho$ and $\omega$ weak NN couplings calculated using nuclear few body techniques~\cite{Viviani2014}. 

Both of these experimental results are in agreement with the Gardner and Muralidhara calculations, which predict $h^1_\pi= 2.14 \pm 0.21 +(+0.17/-0.34) \times 10^{-7}$, to be compared to the value $h^1_\pi= 2.6 \pm 1.2(stat) \pm 0.2 (sys) \times 10^{-7}$ inferred from NPDGamma, and $h_{\rho-\omega}= -12.2  \pm 0.62 +( 0.07/-0.74) \times 10^{-7}$ for the linear combination of NN weak meson couplings $h_{\rho-\omega}=h_{0}^{\rho} + 0.61 h_{0}^{\omega} - 0.61 h_{1}^{\rho} -1.32 h_{1}^{\omega} + 0.03 h_{2}^{\rho} = (-17.0 \pm 6.56) \times 10^{-7}$ which the $A_{PV}$ observable in the $\vec{n}+^{3}$He$\to ^{3}$H$+p$ reaction is sensitive to~\cite{Viviani2014}. In both cases the error estimates for the theory predictions come from the lattice QCD inputs and from the effects in the change in the operator matching conditions of the renormalization group evolution of the NN weak 4-quark operators and the associated couplings over a scale variation of 1.5 GeV (upper case) to 4 GeV (lower case), respectively, around the chosen matching scale of 2 GeV.

This recent theoretical progress allows us to estimate the standard model contribution to $d\phi/dz$  in
n-$^{4}$He. Two calculations of this observable in the meson exchange approach have been published in the scientific literature. The first pioneering calculation~\cite{dmitriev_1983_smr17-0ng02} employed a simplified NN strong interaction potential and reported $d\phi/dz = -0.97 h^1_\pi -0.22 h^0_\omega +0.22
h^1_\omega -0.32h^0_\rho +0.11h^1_\rho$. A subsequent independent calculation performed almost three decades later using 5-body Faddeev-Yakubovsky equations with a modern	EFT treatment of parity-conserving NN interactions~\cite{Lazauskas:2019} reported $d\phi/dz = -0.14 h^1_\pi -0.059 h^0_\omega +0.03 h^1_\omega -0.40h^0_\rho +0.027h^1_\rho$. 
In both cases the $\Delta I=2$ contribution from $h^2_\rho$ was negligible. Green's Function Monte Carlo (GFMC) calculations of parity violation in $n+^{4}$He are also possible now and may be the subject of future work. 

Taking the values of the weak meson exchange couplings from the work of Gardner and Muralidhara mentioned above, one gets a prediction of $d\phi/dz=1.6 \times 10^{-7}$ rad/m from the pioneering Dmitriev et al~\cite{dmitriev_1983_smr17-0ng02} and $d\phi/dz=4.0 \times 10^{-7}$ rad/m from the Lazauskas and Song calculation. Both of these values from standard model parity violation are far enough below the upper bound from experiment of $d\phi/dz =(+2.1 \pm 8.3 (stat.) +2.9/-0.2 (sys.))\times10^{-7} $ rad/m that we can use this data to constrain exotic parity violation.   

To judge whether or not the upper bound on $A_{L}$ in $n+^{232}$Th quoted above might possess contributions from standard model parity violation near the experimental upper bound, one must extend the meson exchange model of NN weak interactions at the nucleon-nucleon level to the case of neutron-nucleus interactions. Fortunately this theoretical work has been performed in the past and described in several works~\cite{Zaretsky1983, Flambaum1984, Noguera1986, Flambaum1992, Flambaum1993DynamicalEO} which we summarize and describe here. In the nonrelativistic limit one can write the effective Hamiltonian for P-odd neutron interactions with a nucleus in a form first proposed by Michel~\cite{Michel:1964} and employed in later work~\cite{Karl1977}  

\begin{equation}
H=p^{2}/2m +V(r)+{1 \over 2} \{ F(r), {\vec{\sigma} \cdot \vec{p}}  \} 
\end{equation}

where $\sigma$ is the neutron spin, $\vec{p}$ is the neutron momentum, and $F(r)$ is the weak potential. The weak interaction Hamiltonian is given by the anti-commutator between the weak potential, $F(r)$, and the neutron helicity operator $\vec{\sigma}\cdot\vec{p}$ 

$$ H_{weak}=\{F(r),\vec{\sigma}\cdot\vec{p}\}=F(r)\vec{\sigma}\cdot\vec{p}-iF_r(r)\vec{\sigma}\cdot\hat{r}$$

where the subscript $r$ indicates the first derivative with respect to r. The parity-odd operator $\vec{\sigma}\cdot\vec{p} $ in the first term is the one which leads to both the helicity dependence of the total cross section $P$ and the parity-odd neutron rotary power $d\phi/dz$. The $\sigma\cdot\hat{r}$ term gives a subdominant contribution when the matrix element of the Hamiltonian between plane wave states is taken to calculate the forward scattering amplitude~\cite{Karl1977, Forte1978} and will be neglected in the rest of this paper. 

The weak potential $F(r)$ is fixed by writing the
neutron-nucleus weak interaction in a form that explicitly includes the nuclear density,
\begin{align}
  V_{\rm PNC} &= \varepsilon g_{n}\,
     \{\bm{\sigma}\cdot\mathbf{p},\,\rho(r)\},
  \nonumber \\
  \rho(r) &= \frac{\rho_N(r)}{\rho_0},
  \qquad \rho_0 = 0.148~\mathrm{fm}^{-3},
  \label{eq:VPNC}
\end{align}
so that $F(r) = \varepsilon g_{n}\rho$. $\varepsilon g_{n}$ parametrizes the strength of the parity-odd neutron-nucleus interaction. For potential scattering the parity-odd forward amplitude generated by Eq.~(\ref{eq:VPNC}) is~\cite{Mitchell:2000}

\begin{equation}
  f' = 4\varepsilon g_{n}\,\frac{m_n}{k}\,f_p
       \left(1 + \frac{f_s}{2R}\right),
  \label{eq:fprime}
\end{equation}
where $f_s$ and $f_p$ are the strong $s$- and $p$-wave amplitudes and the second term in the
bracket comes from the surface term. In the low-energy limit the nuclear radius enters through the $p$-wave hard-sphere amplitude and the surface term, $f_p \simeq -(kR)^2R/3$, while for the $s$-wave amplitude one can use the measured potential scattering length, $f_s = -b_c$ to give

\begin{align}
  f' &= -\frac{4}{3}\left(1-\frac{b_c}{2R}\right)
        \varepsilon g_{n}\, m_n k R^{3},
  \nonumber \\
  \frac{\Delta\sigma}{2\sigma}
     &= \frac{8}{3}\left(1-\frac{b_c}{2R}\right)
        \frac{\varepsilon g_{n}\, m_n k R^{3}}{b_c},
  \label{eq:Pofflead}
\end{align}

We use the same notation as in~\cite{Mitchell:2000}, who show that this expression matches that derived in several previous works~\cite{Zaretsky1983, Flambaum1984, Noguera1986, Flambaum1992, Flambaum1993DynamicalEO}. Our information on the weak neutron-nucleus coupling $g_{n}$ has recently improved. The result of $A^{np}_{\gamma}=[-3.0 \pm 1.4(stat) \pm 0.2(sys)] \times 10^{-8}$ for the P-odd asymmetry in $\vec{n}+p \to d+\gamma$ reported by the NPDGamma collaboration~\cite{NPDGamma:2018vhh} enabled Flambaum and Mansour~\cite{PhysRevC.105.015501} to update older calculations of $g_{n}$ based on Fermi liquid theory applied to the weak NN meson exchange model using DDH best values as input from the value $g_{n}=1.0$ used in previous work~\cite{Flambaum1984, Flambaum1992, Flambaum1993DynamicalEO} to $g_{n}=0.9$. So now the NN weak meson exchange model value of $\varepsilon g_{n}$ is on a more secure, experimentally-based foundation. The theoretical precision of the Fermi liquid-based calculation of $\varepsilon g_{n}$ at the 50\% level~\cite{Flambaum1993DynamicalEO} is accurate enough for our purposes.     

The expression for $\Delta \sigma$ is consistent with the threshold suppression of $A_{L}$: for slow neutron--nucleus scattering in the regime $kR \ll 1$ and in the absence of neutron-nucleus resonances, the imaginary part of the parity-odd forward amplitude is small compared with the real part.

\begin{equation}
  \frac{Im(\Delta F)}{Re(\Delta F)}= k\,Re f_0(E) \simeq k b_c
  \label{eq:eta}
\end{equation}

where $b_{c}$ is the coherent s-wave neutron-nucleus scattering length. This follows when one writes the scattering amplitude matrix within the $K$-matrix formalism~\cite{Stodolsky}, which explicitly takes unitarity into account as $f = k^{-1}K(1-iK)^{-1}$ where $k$ is the neutron momentum and $K$ Hermitian leads to a unitary $f$. As $k \to 0$ in the $kR<<1$ limit, the threshold behavior of $K \to k_{f}^{l_{f}+1/2}Mk_{i}^{l_{i}+1/2}$ where $M$ is a slowly-varying amplitude at threshold, $k_{i}$ and $k_{f}$ are the incident and final momenta in the partial wave reaction channels with orbital angular momenta $l_{i}$ and $l_{f}$. Since $K$ vanishes at threshold and the parity-violating matrix element requires $f$ to connect $s$ to $p$ waves, one gets $Re(f) = K/k \to constant$ and $Im(f) = KK/k \propto k$. Since this suppression is kinematic in nature and depends only on the P-odd property of the amplitude, this same suppression factor applies also to any exotic P-odd neutron-nucleus interaction contribution to $P$ as well. As we will see below this will make the constraints from $A_{L}$ in $n+^{232}$Th weaker than those from $d\phi \over dz$ in $n+^{4}$He.

For $^{232}$Th we have $A = 232$, $Z = 90$ and $N = 142$. If we take the nuclear radius in the form $R = r_0 A^{1/3}$ with $r_0 = 1.35$~fm as used in previous work, then $R_{\rm Th} = 1.35\,A^{1/3} = 8.30~\mathrm{fm}$ and $k R_{\rm Th} = 1.82\times 10^{-3}\,\sqrt{E/\mathrm{eV}}$ and $kR_{\rm Th} = 0.0100$--$0.0316$ across the analyzed neutron energy band of $30$--$300$~eV. For quantitative evaluation it is better to use the measured $n-^{232}$Th bound coherent scattering length

\begin{align}
  b_c &= 10.31(3)~\mathrm{fm},
  \nonumber \\
  \sigma_{\rm coh} = \sigma_{\rm scatt} &= 13.36(8)~\mathrm{b},
  \nonumber \\
  \sigma_{\rm abs} &= 7.37(6)~\mathrm{b},
  \label{eq:bc}
\end{align}

where $\sigma_{\rm abs}$ is quoted at the conventional thermal neutron energy of $25$~meV. Numerically one gets

\begin{align}
  \frac{\Delta\sigma}{2\sigma}
     &= 0.0585\,\varepsilon g_{n}\,\sqrt{E/\mathrm{eV}},
  \nonumber \\
  |\epsilon_{\rm pnc}(E)|
     &= 0.191\,|\varepsilon g_{n}|\,\sqrt{E/\mathrm{eV}},
  \label{eq:coefficients}
\end{align}
where the second line follows from Eq.~(\ref{eq:epspnc}). These are the two
coefficients the data are fitted against. $\sigma_{\rm abs}$ scaled by the $1/v$ law gives $0.21$~b at $30$~eV and $0.068$~b at $300$~eV, that is $1.6\%$ and $0.5\%$ of the scattering cross
section, so $\sigma_+ + \sigma_- \simeq 2\sigma_{\rm scatt}$ holds to better than $2\%$. For the target areal density $N_A = 3.40\times 10^{-3}~\mathrm{fm}^{-2}$ used in the experiment and neutron polarization $f_n = 0.72$, $N_A f_n \sigma = 3.27$. The measured transmission asymmetry is related to the longitudinal cross-section asymmetry

\begin{equation}
  P_{\rm off} = \frac{\sigma_+ - \sigma_-}{\sigma_+ + \sigma_-}
              \simeq \frac{\Delta\sigma}{2\sigma},
  \qquad \Delta\sigma = \sigma_+ - \sigma_- ,
  \label{eq:poffdef}
\end{equation}
by 
\begin{align}
  \epsilon_{\rm pnc} &\simeq -\tfrac{1}{2}\,N_A f_n \Delta\sigma
                      = -N_A f_n \sigma P_{\rm off},
  \nonumber \\
  |\epsilon_{\rm pnc}| &= 3.27\,|P_{\rm off}| .
  \label{eq:epspnc}
\end{align}

The theoretical estimate of $P_{off, SM}={{{\sigma_{+}-\sigma_{-}}\over{{\sigma_{+}-\sigma_{-}}}}}\ \leq 0.4 \times 10^{-6}$ presented in Mitchell et al made the most optimistically large possible assumption for the size of the neutron-nucleus weak interaction amplitude. Evaluation of the expressions above, which are now constrained by the data from NPDGamma, now gives $\varepsilon_{\rm SM} = 1.00\times 10^{-8}\,g_n$ with $g_n=0.9$ the dimensionless neutron-nucleon weak coupling. Now $P_{off} \leq 0.3 \times 10^{-7}$, a factor of 50 smaller than the experimental upper bound. Furthermore the standard model contribution from the parity-odd neutron-electron interaction is smaller still as it is suppressed relative to the neutron-nucleon weak interaction by a factor of $1-4 \sin^{2} (\theta_{W}) \approx 0.1$~\cite{Tadic1978}. We therefore conclude that the Standard Model parity violation background for the off-resonance $n+^{232}$Th data is negligible so that we can interpret our upper bound solely in terms of a limit on exotic parity-odd amplitudes.

Mitchell et al applied an energy-independent model for the P-odd asymmetry to the twenty energy intervals between the resonances which they analyzed to obtain $\epsilon_{\rm pnc} = (0.162 \pm 0.520)\times 10^{-5}$ with $\chi^2_\nu = 0.87$. If one takes into account the $\sqrt{E}$ dependence of a parity-odd amplitude from neutron optical kinematics, one should use a model  
\begin{equation}
  \overline{\sqrt{E}} = \frac{2}{3}\,
      \frac{E_f^{3/2} - E_i^{3/2}}{E_f - E_i},
  \label{eq:intervalmean}
\end{equation}

over the intervals $E_i < E < E_f$. With $\epsilon^{(j)}_{\rm pnc} = X_j\varepsilon$ and
$X_j = 0.191\,\overline{\sqrt{E}}_j$ the weighted estimator and its uncertainty are
\begin{align}
  \hat\varepsilon &= \frac{\sum_j X_j y_j/\sigma_j^2}
                          {\sum_j X_j^2/\sigma_j^2},
  \nonumber \\
  \sigma_\varepsilon &= \Big(\textstyle\sum_j X_j^2/\sigma_j^2\Big)^{-1/2}.
  \label{eq:estimator}
\end{align}

\begin{align}
  \varepsilon &= (1.32 \pm 2.53)\times 10^{-6},
  \qquad \chi^2_\nu = 0.87,
  \nonumber \\
  |\varepsilon|_{95} &= |\hat\varepsilon| + 1.96\,\sigma_\varepsilon
                      = 6.3\times 10^{-6},
  \label{eq:eps95}
\end{align}
with $\nu = 19$, equivalently
$\Delta\sigma/2\sigma = (0.77 \pm 1.48)\times 10^{-7}\sqrt{E/\mathrm{eV}}$. The second line only differs slightly from the Mitchell et al result quoted above.

\section{Long-range exotic interaction}
\label{sec:longrange}
 
\subsection{Forward Born amplitude}

The axial-vector/vector potential between a polarized neutron and a nucleon,
in the convention of Refs.\cite{fadeev_revisiting_2019,Cong_rev}, is
\begin{equation}
  V_{12+13} = \frac{g^n_A g^N_V}{8\pi m_n}
    \left\{\bm{\sigma}\cdot\mathbf{p},\;\frac{e^{-r/\lambda}}{r}\right\},
  \label{eq:yukawa}
\end{equation}
where $\lambda$ is the range of the exchange boson. We use throughout the per-nucleon vector coupling
\begin{equation}
  g^N_V = \frac{N g^n_V + Z g^p_V}{A} .
  \label{eq:gVN}
\end{equation}
 For an unpolarized source Eq.~(\ref{eq:yukawa}) possesses a spin-velocity form
$V_{\rm PV}(r) = (g/4\pi)(e^{-r/\lambda}/r)\,\bm{\sigma}\cdot\mathbf{v}$ with
$g \equiv g^n_A g^N_V$. Applying the Born approximation to calculate the scattering amplitude gives

\begin{align}
  \Delta f_N \equiv f^{\rm PV}_+ - f^{\rm PV}_-
    &= -\frac{g k \lambda^2}{\pi},
  \label{eq:born}
\end{align}
and for $\lambda \gg R_{\rm Th}$ the $A$ nucleons add coherently so that $\Delta F_A = A\,\Delta f_N$. For an
unpolarized target the relations of Ref.~\cite{Stodolsky} give 

\begin{equation}
  \frac{\phi}{L} = \frac{2\pi}{k}\,\rho\,Re(\Delta F),
  \qquad
  \Delta\sigma = \frac{4\pi}{k}\,Im(\Delta F),
  \label{eq:stodolsky}
\end{equation}
so that the long-range spin rotation is
$d\phi/dL = -2 \rho g \lambda^2$ and, for a finite target,
\begin{equation}
  |\phi| = 2 \rho L |g| \lambda^2 .
  \label{eq:rotation}
\end{equation}

From the transmission definition and Eqs.~(\ref{eq:stodolsky}), (\ref{eq:rotation}), and (\ref{eq:eta}),

\begin{align}
  |\epsilon_{\rm pnc}(E)| &= 2 f_n N_N |g| \lambda^2\, k(E)\, b_c,
  \nonumber \\
  2 f_n N_A b_c &= 11.71~\mathrm{fm},
  \label{eq:lrcoeff}
\end{align}
and with $G \equiv g^n_A g^N_V \lambda^2$,
\begin{equation}
  |\epsilon_{\rm pnc}(E)| = 2.573\times 10^{-3}\,
    \sqrt{E/\mathrm{eV}}\;\big|G/\mathrm{fm}^2\big| .
  \label{eq:lrmodel}
\end{equation}
 
\subsection{Experimental Constraints}

The result for $n+^{232}$Th from the analysis presented above is
\begin{align}
  G &= (0.91 \pm 1.95)\times 10^{-4}~\mathrm{fm}^2,
  \qquad \chi^2_\nu = 0.87,
  \nonumber \\
  |g^n_A g^N_V|\lambda^2 &< 4.7\times 10^{-4}~\mathrm{fm}^2,
  \qquad \lambda \gg R_{\rm Th}.
  \label{eq:Clong}
\end{align}

and for $n+^{4}$He we have

\begin{equation}
  |g^n_A g^N_V|_{{}^4{\rm He}} <
    \frac{5.0\times 10^{-5}}{\lambda^2[\mathrm{fm}^2]} .
  \label{eq:CHe}
\end{equation}

\section{Comparison of the constraints}
\label{sec:compare}
 
\begin{figure*}[t]
  \includegraphics[width=0.86\textwidth]{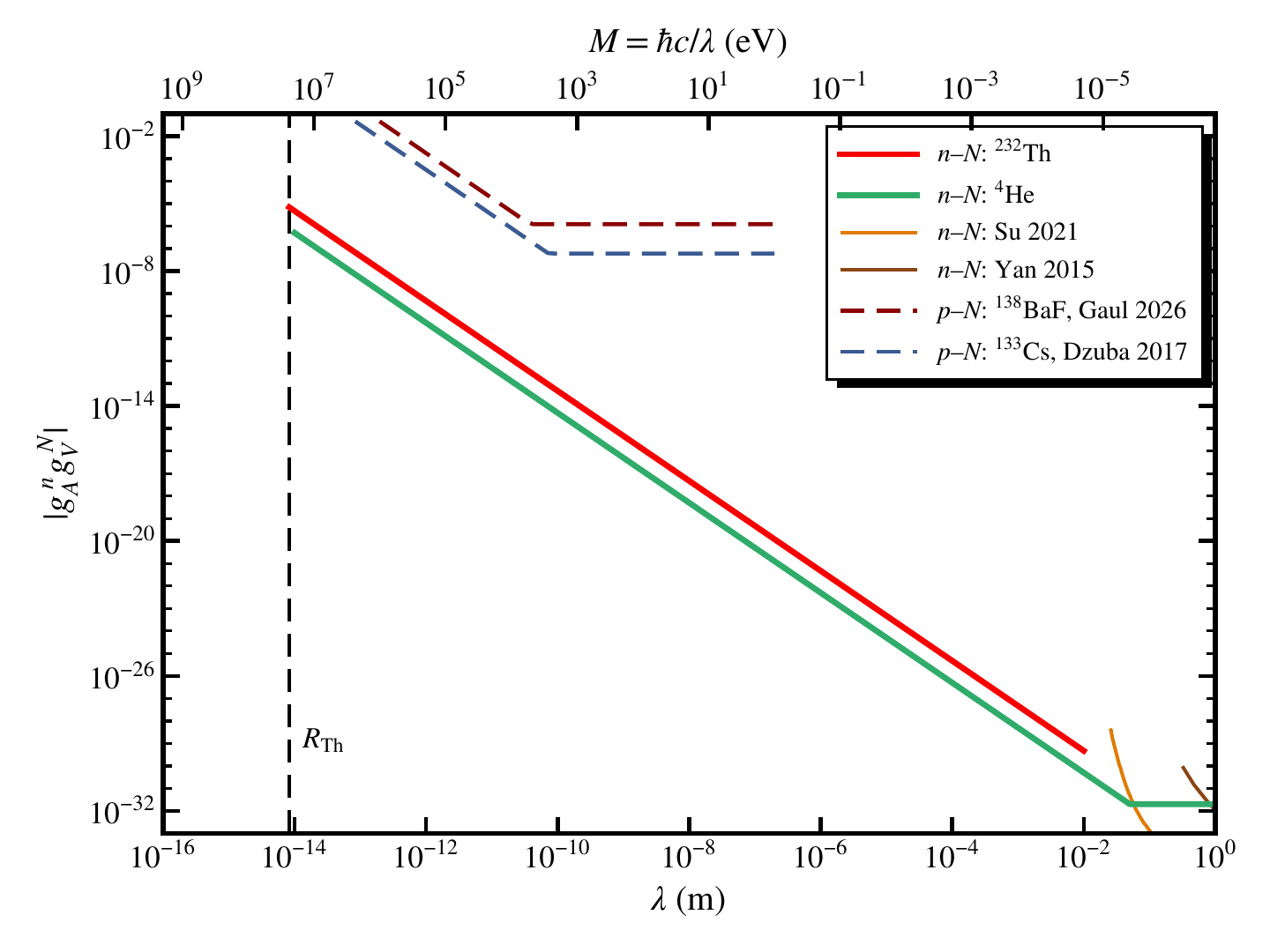}
  \caption{Constraints on the coupling constant product $|g^n_A g^N_V|$ as a function of the exchange boson range $\lambda$. The $^{232}$Th and $n$-$^4$He constraints 
  are given in Eqs.~(\ref{eq:Clong}) and (\ref{eq:CHe}). 
  The flattening above $\lambda \approx 5$~cm of the $n$--$^4$He constraint marks where finite target geometry
  stops the $\lambda^2$ growth. The thin lines show other direct
  experimental limits: $n$--$N$ limits from a spin-based amplifier~\cite{su_search_2021}
  and from spin relaxation of polarized $^3$He~\cite{yan_searching_2015}, taken
  from the compilation of Ref.~\cite{Cong_rev}. The dashed lines are limits on $|g^p_A g^N_V|$ from parity violation in $^{138}$BaF~\cite{Altuntas2018,Gaul2026}
  and in $^{133}$Cs~\cite{Dzuba2017}, taken from Ref.~\cite{Gaul2026}. The
  vertical marker is $R_{\rm Th}$.}

  \label{fig:constraints}
\end{figure*}

Figure~\ref{fig:constraints} compares our results with other direct
experimental limits on exotic axial-vector/vector interactions. From
$R_{\rm Th}$ up to $\lambda \approx 6$~cm the $^{232}$Th and $n$--$^4$He
results are the strongest direct experimental limits on $g^n_A g^N_V$; the
other direct limits on this product~\cite{su_search_2021,yan_searching_2015,clayburn_using_2023,wu_new_2023}
apply at longer ranges, as do those on $g^p_A g^N_V$ from
Ref.~\cite{shu_constraint_2024}. The $p$--$N$ limits from parity violation in
$^{138}$BaF and $^{133}$Cs~\cite{Gaul2026,Dzuba2017} are weaker than ours by
about six orders of magnitude at $\lambda \sim 20$~fm and by up to twelve
orders of magnitude at longer ranges. 

Stronger bounds in this range of
$\lambda$ can be obtained by combining laboratory limits on the vector
coupling with astrophysical limits on the axial coupling; these
astrophysically assisted bounds depend on additional model assumptions and are
discussed in Ref.~\cite{Cong_rev}. For macroscopic forces ranges of 1 cm and above one can derive much more stringent upper bounds on parity-odd neutron interactions with matter by combining separate laboratory limits on $g_{V}^{2}$ from torsion balance experiments and from $g_{A}^{2}$ from atomic physics measurements~\cite{macro3}.

\begin{figure*}[t]
  \includegraphics[width=0.86\textwidth]{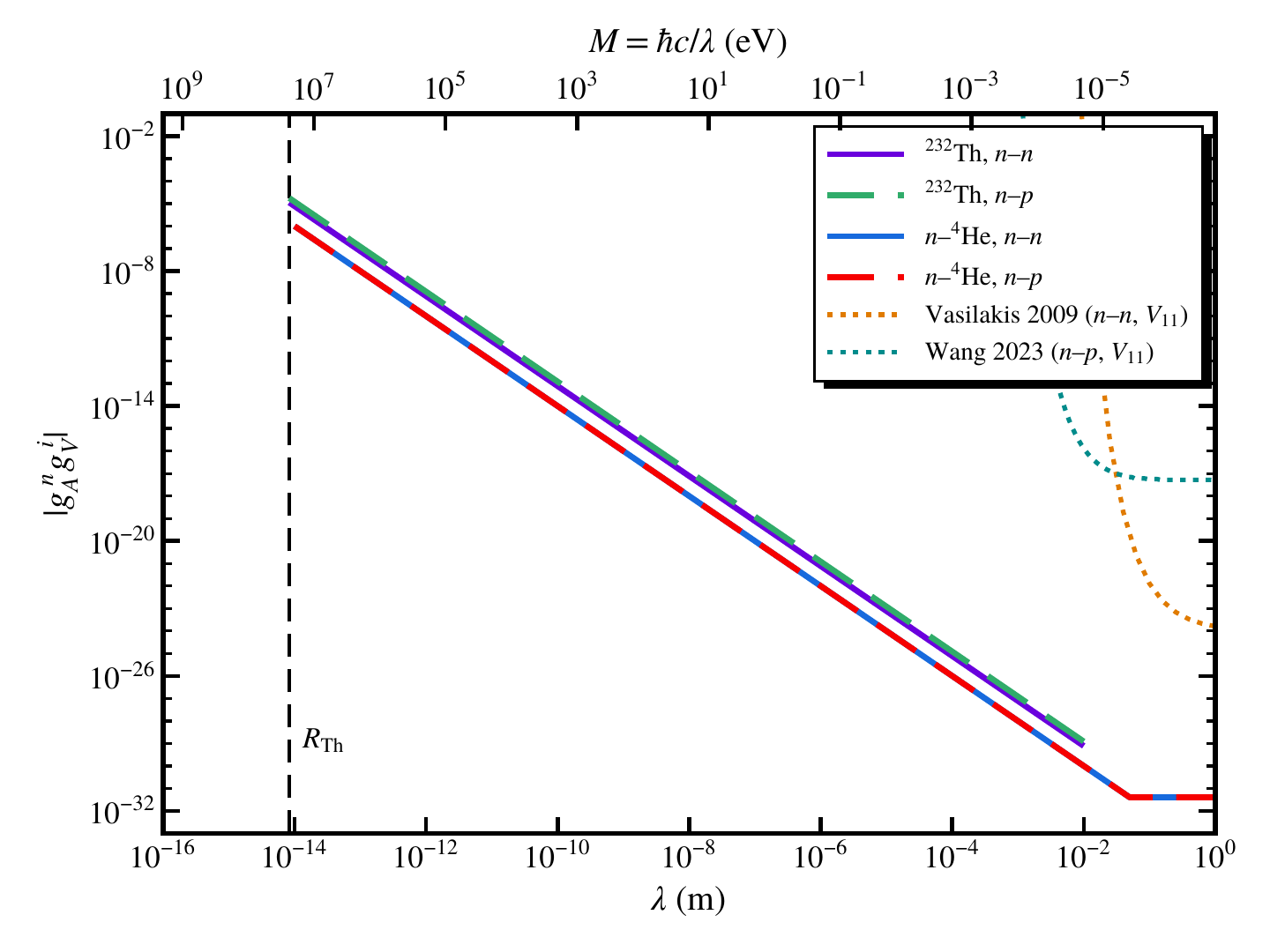}
  \caption{Limits on neutron-neutron and neutron-proton couplings assuming the other vanishes. The $^{232}$Th
  and $n$--$^4$He results are shown over the same ranges as in
  Fig.~\ref{fig:constraints} as given in Eq.~(\ref{eq:oneatatime}). The dotted
  lines show the $n$--$n$ limit of Ref.~\cite{vasilakis_limits_2009} and the
  $n$--$p$ limit of Ref.~\cite{wang_search_2023}, both taken from the compilation of Ref.~\cite{Cong_rev}.
  Dropping the one-at-a-time assumption by inverting the two
  measurements jointly gives the weaker limits of Eq.~(\ref{eq:joint}).}
  \label{fig:separated}
\end{figure*}

The coherent forward amplitude sums over both the nucleons and the electrons in the atom. An electron enters with a coupling $g^e_V$ in place of $g^N_V$, so $g^p_V$ and $g^e_V$ appear in each observable only through their sum $S = N g^n_V + Z\big(g^p_V + g^e_V\big)$

Writing $X \equiv g^n_A g^n_V \lambda^2$ and
$Y \equiv g^n_A (g^p_V + g^e_V)\lambda^2$, each measurement constrains a weighted combination,
\begin{align}
  \left|\frac{N}{A}X + \frac{Z}{A}Y\right| &< C_{\rm Th},
  \nonumber \\
  \left|\tfrac{1}{2}X + \tfrac{1}{2}Y\right| &< C_{{}^4{\rm He}},
  \label{eq:system}
\end{align}
with $N/A = 0.61$ and $Z/A = 0.39$ for $^{232}$Th, $N = Z = 2$ for $^{4}$He. If we assume the exchange boson couples to one nucleon species at a time we get limits on the surviving coupling of $|X| < C/(N/A)$ and $|Y| < C/(Z/A)$ shown in Fig.~\ref{fig:separated}, which in fm$^2$ are

\begin{align}
  n\mathrm{-}^{232}\mathrm{Th:}&\quad
    |X| < 7.7\times 10^{-4}, \;\; |Y| < 1.2\times 10^{-3},
  \nonumber \\
  n\mathrm{-}^4\mathrm{He:}&\quad
    |X| < 1.0\times 10^{-4}, \;\; |Y| < 1.0\times 10^{-4}.
  \label{eq:oneatatime}
\end{align}

Using the two measurements together, inverting the
$2\times 2$ system of Eq.~(\ref{eq:system}) and bounding each combination by the triangle inequality gives

\begin{align}
  |X| &< 4.46\,C_{\rm Th} + 3.46\,C_{{}^4{\rm He}}
       = 2.3\times 10^{-3},
  \nonumber \\
  |Y| &< 4.46\,C_{\rm Th} + 5.46\,C_{{}^4{\rm He}}
       = 2.4\times 10^{-3}~\mathrm{fm}^2,
  \label{eq:joint}
\end{align}
weaker by about a factor of 3 than the one-at-a-time bounds. 

The separated limits of Fig.~\ref{fig:separated} are stronger than the only
other limits that distinguish $n$--$n$ and $n$--$p$
couplings~\cite{vasilakis_limits_2009,wang_search_2023} by many orders of
magnitude, and they extend to much shorter ranges.

\section{Quantum dark forces}
\label{sec:qdf}

It is worth considering whether or not these experiments can constrain other possible sources of parity-odd neutron-nucleon interactions. Brax, Fichet, and Pignol~\cite{PhysRevLett.120.131801, PhysRevD.97.115034} considered and analyzed the exchange of light (sub-GeV) dark sector particles of mass $m$ which couple bilinearly to standard model particles rather than linearly. The authors showed that these interactions could be analyzed in a general way using an effective field theory approach valid for energy scales below $\Lambda_{QCD}$ and applied to operators of the form $O_{nuc}O_{DS}$, where $O_{DS}$ is a dark sector field bilinear and $O_{nuc}$ is a fermion bilinear of the form $\bar{N}\Gamma N$ with $N$ a fermion field and $\Gamma$ an arbitrary Lorentz structure. After Brax, Fichet, and Pignol analyzed data from several spin-independent experiments to constrain such interactions, their results were extended to the case of spin-dependent quantum dark forces between nucleons~\cite{Costantino:2019ixl}.

We considered whether or not we can use the neutron optical experiments analyzed here to constrain dark fermion exchange, which can involve both vector and axial-vector channels and therefore can violate parity. The coupling of a spin-$1/2$ dark fermion $\chi$ to the nucleon current
\begin{equation}
  \mathcal{L}_{\rm eff} = \frac{1}{\Lambda^{2}}\,
    \bar N \gamma^{\mu}\!\left(c_V + c_A\gamma^{5}\right)\! N \;
    \bar\chi\gamma_{\mu}\chi .
  \label{eq:Leff}
\end{equation}

for $c_V = -c_A$ generates the maximally parity-violating left-handed combination
$\bar N\gamma^{\mu}(1-\gamma^{5})N\,\bar\chi\gamma_{\mu}\chi$. The nucleon--nucleon potential generated by a
$\chi\bar\chi$ pair also produces parity-conserving pieces $V\!\otimes\!V$ and an $A\!\otimes\!A $ together with the parity-odd interference term proportional to $c_Vc_A$. In the nonrelativistic limit the vector vertex supplies the
nucleon density and the axial vertex the neutron helicity operator, so that

\begin{equation}
  V^{\rm QDF}_{\rm PV}(\mathbf{r})
    = \frac{c_Vc_A}{2m_n}
      \Big\{\bm{\sigma}\cdot\mathbf{p},\;\mathcal{V}(r)\Big\},
  \label{eq:VPVQDF}
\end{equation}
with $\mathcal{V}(r)$ the relevant potential, just like the Michel
form of Eq.~(\ref{eq:VPNC}). The resulting potential is

\begin{equation}
  \mathcal{V}(r) = \frac{\mathcal{C}}{\Lambda^{4}}\,\frac{1}{r^{5}},
  \label{eq:profile}
\end{equation}

with $\mathcal{C}$ a dimensionless coefficient fixed by the loop calculation. This structure is familiar from the very similar calculation of neutrino pair
exchange~\cite{PhysRev.166.1638}, to which Eq.~(\ref{eq:Leff}) reduces on
replacing $\Lambda^{-2}$ by $G_F/\sqrt{2}$ and $\chi$ by a neutrino. For the forward scattering amplitude involved in the experiments discussed in this work, this term yields an amplitude proportional to 

\begin{align}
  \int d^{3}r\;\frac{1}{r^{5}} &= \frac{2\pi}{r_{\rm min}^{2}} .
  \label{eq:contrast}
\end{align}

whose value is very sensitive to the poorly-known short-range details of the nucleon-nucleon interactions in the nucleus and is therefore outside the range of validity of our present analysis.

\section{Conclusion}

We have presented new limits on exotic parity-odd interactions between neutrons and nucleons from sensitive experimental upper bounds on parity violation in epithermal neutron-nucleus total cross section in $^{232}$Th in the eV-keV neutron energy range and in the parity-odd rotary power in $n+^{4}$He in the meV neutron energy range. We used recent progress in our understanding of the weak interaction between nucleons to confirm that the Standard Model expectation for both the $n+^{4}$He and the $n+^{232}$Th experimental results fall far enough below the present upper limits to conduct our analysis. The large asymmetry between neutron and proton numbers in the $^{232}$Th nucleus allows us to constrain separately the neutron-neutron, neutron-proton, and neutron-electron vector and axial vector coupling constant product combinations $g^n_A g^n_V$ and $g^n_A (g^p_V + g^e_V)$ for the first time.

Additional progress on this question would require an improved understanding of the Standard Model weak NN interaction which is not likely to be forthcoming in the near future. One would need to conduct sensitive NN weak interaction measurements in two body and few body systems to add more data and enable theoretical calculations of the NN weak interaction to be tested before one could confidently correct for standard model effects to reveal exotic parity violation.  

New experimental approaches may also improve the sensitivity to these
interactions. Using molecular electronic-structure calculations,
Gaul, Cong and Budker~\cite{Gaul2026} have shown that parity-violation
measurements with heavy polar diatomic molecules such as $^{137}$BaF,
$^{225}$RaF and $^{29}$SiO$^+$ could improve the present sensitivity to
axial-vector/vector nucleon--nucleus interactions by up to several orders of
magnitude, including sensitivity to the neutron coupling through nuclei with an
unpaired neutron. Such data and calculations can also separately constrain the product $g^n_A g^e_V$. If this idea can be realized, our bounds could be used in combination with such results to derive separate constraints on $g^n_A g^p_V$. 

\begin{acknowledgments}
This research was funded by the Science Committee of the Ministry of Science and Higher Education of the Republic of Kazakhstan (Grant No. AP23484023). W. M. Snow acknowledges support from US National Science Foundation grant PHY-2209481 and from the Indiana University Center for Spacetime Symmetries.
\end{acknowledgments}


\begin{thebibliography}{91}%
\makeatletter
\providecommand \@ifxundefined [1]{%
 \@ifx{#1\undefined}
}%
\providecommand \@ifnum [1]{%
 \ifnum #1\expandafter \@firstoftwo
 \else \expandafter \@secondoftwo
 \fi
}%
\providecommand \@ifx [1]{%
 \ifx #1\expandafter \@firstoftwo
 \else \expandafter \@secondoftwo
 \fi
}%
\providecommand \natexlab [1]{#1}%
\providecommand \enquote  [1]{``#1''}%
\providecommand \bibnamefont  [1]{#1}%
\providecommand \bibfnamefont [1]{#1}%
\providecommand \citenamefont [1]{#1}%
\providecommand \href@noop [0]{\@secondoftwo}%
\providecommand \href [0]{\begingroup \@sanitize@url \@href}%
\providecommand \@href[1]{\@@startlink{#1}\@@href}%
\providecommand \@@href[1]{\endgroup#1\@@endlink}%
\providecommand \@sanitize@url [0]{\catcode `\\12\catcode `\$12\catcode `\&12\catcode `\#12\catcode `\^12\catcode `\_12\catcode `\%12\relax}%
\providecommand \@@startlink[1]{}%
\providecommand \@@endlink[0]{}%
\providecommand \url  [0]{\begingroup\@sanitize@url \@url }%
\providecommand \@url [1]{\endgroup\@href {#1}{\urlprefix }}%
\providecommand \urlprefix  [0]{URL }%
\providecommand \Eprint [0]{\href }%
\providecommand \doibase [0]{https://doi.org/}%
\providecommand \selectlanguage [0]{\@gobble}%
\providecommand \bibinfo  [0]{\@secondoftwo}%
\providecommand \bibfield  [0]{\@secondoftwo}%
\providecommand \translation [1]{[#1]}%
\providecommand \BibitemOpen [0]{}%
\providecommand \bibitemStop [0]{}%
\providecommand \bibitemNoStop [0]{.\EOS\space}%
\providecommand \EOS [0]{\spacefactor3000\relax}%
\providecommand \BibitemShut  [1]{\csname bibitem#1\endcsname}%
\let\auto@bib@innerbib\@empty
\bibitem [{\citenamefont {Leitner}\ and\ \citenamefont {Okubo}(1964)}]{Leitner64}%
  \BibitemOpen
  \bibfield  {author} {\bibinfo {author} {\bibfnamefont {J.}~\bibnamefont {Leitner}}\ and\ \bibinfo {author} {\bibfnamefont {S.}~\bibnamefont {Okubo}},\ }\bibfield  {title} {\bibinfo {title} {Parity, charge conjugation, and time reversal in the gravitational interaction},\ }\href {https://doi.org/10.1103/PhysRev.136.B1542} {\bibfield  {journal} {\bibinfo  {journal} {Phys. Rev.}\ }\textbf {\bibinfo {volume} {136}},\ \bibinfo {pages} {B1542} (\bibinfo {year} {1964})}\BibitemShut {NoStop}%
\bibitem [{\citenamefont {Weinberg}(1972)}]{Weinberg72}%
  \BibitemOpen
  \bibfield  {author} {\bibinfo {author} {\bibfnamefont {S.}~\bibnamefont {Weinberg}},\ }\bibfield  {title} {\bibinfo {title} {Approximate symmetries and pseudo-goldstone bosons},\ }\href {https://doi.org/10.1103/PhysRevLett.29.1698} {\bibfield  {journal} {\bibinfo  {journal} {Phys. Rev. Lett.}\ }\textbf {\bibinfo {volume} {29}},\ \bibinfo {pages} {1698} (\bibinfo {year} {1972})}\BibitemShut {NoStop}%
\bibitem [{\citenamefont {Moody}\ and\ \citenamefont {Wilczek}(1984)}]{Moody84}%
  \BibitemOpen
  \bibfield  {author} {\bibinfo {author} {\bibfnamefont {J.~E.}\ \bibnamefont {Moody}}\ and\ \bibinfo {author} {\bibfnamefont {F.}~\bibnamefont {Wilczek}},\ }\bibfield  {title} {\bibinfo {title} {New macroscopic forces?},\ }\href {https://doi.org/10.1103/PhysRevD.30.130} {\bibfield  {journal} {\bibinfo  {journal} {Phys. Rev. D}\ }\textbf {\bibinfo {volume} {30}},\ \bibinfo {pages} {130} (\bibinfo {year} {1984})}\BibitemShut {NoStop}%
\bibitem [{\citenamefont {Zwiebach}(2004)}]{string1}%
  \BibitemOpen
  \bibfield  {author} {\bibinfo {author} {\bibfnamefont {B.}~\bibnamefont {Zwiebach}},\ }\href@noop {} {\emph {\bibinfo {title} {A First Course in String Theory}}}\ (\bibinfo  {publisher} {Cambridge University Press},\ \bibinfo {year} {2004})\BibitemShut {NoStop}%
\bibitem [{\citenamefont {Green}\ \emph {et~al.}(2012{\natexlab{a}})\citenamefont {Green}, \citenamefont {Schwarz},\ and\ \citenamefont {Witten}}]{string2}%
  \BibitemOpen
  \bibfield  {author} {\bibinfo {author} {\bibfnamefont {M.~B.}\ \bibnamefont {Green}}, \bibinfo {author} {\bibfnamefont {J.~H.}\ \bibnamefont {Schwarz}},\ and\ \bibinfo {author} {\bibfnamefont {E.}~\bibnamefont {Witten}},\ }\href@noop {} {\emph {\bibinfo {title} {Superstring Theory: 25th Anniversary Edition}}}\ (\bibinfo  {publisher} {Cambridge University Press},\ \bibinfo {year} {2012})\BibitemShut {NoStop}%
\bibitem [{\citenamefont {Green}\ \emph {et~al.}(2012{\natexlab{b}})\citenamefont {Green}, \citenamefont {Schwarz},\ and\ \citenamefont {Witten}}]{string3}%
  \BibitemOpen
  \bibfield  {author} {\bibinfo {author} {\bibfnamefont {M.~B.}\ \bibnamefont {Green}}, \bibinfo {author} {\bibfnamefont {J.~H.}\ \bibnamefont {Schwarz}},\ and\ \bibinfo {author} {\bibfnamefont {E.}~\bibnamefont {Witten}},\ }\href@noop {} {\emph {\bibinfo {title} {Superstring Theory: Volume 2, Loop Amplitudes, Anomalies and Phenomenology: 25th Anniversary Edition}}}\ (\bibinfo  {publisher} {Cambridge University Press},\ \bibinfo {year} {2012})\BibitemShut {NoStop}%
\bibitem [{\citenamefont {Dobrescu}(2005)}]{paraphoton1}%
  \BibitemOpen
  \bibfield  {author} {\bibinfo {author} {\bibfnamefont {B.~A.}\ \bibnamefont {Dobrescu}},\ }\bibfield  {title} {\bibinfo {title} {Massless gauge bosons other than the photon},\ }\href {https://doi.org/10.1103/PhysRevLett.94.151802} {\bibfield  {journal} {\bibinfo  {journal} {Phys. Rev. Lett.}\ }\textbf {\bibinfo {volume} {94}},\ \bibinfo {pages} {151802} (\bibinfo {year} {2005})}\BibitemShut {NoStop}%
\bibitem [{\citenamefont {Holdom}(1986)}]{paraphoton2}%
  \BibitemOpen
  \bibfield  {author} {\bibinfo {author} {\bibfnamefont {B.}~\bibnamefont {Holdom}},\ }\bibfield  {title} {\bibinfo {title} {Two u(1)'s and the $\epsilon$ charge shifts},\ }\href {https://doi.org/10.1016/0370-2693(86)91377-8} {\bibfield  {journal} {\bibinfo  {journal} {Phys. Lett. B}\ }\textbf {\bibinfo {volume} {166}},\ \bibinfo {pages} {196} (\bibinfo {year} {1986})}\BibitemShut {NoStop}%
\bibitem [{\citenamefont {Arvanitaki}\ \emph {et~al.}(2010)\citenamefont {Arvanitaki}, \citenamefont {Dimopoulos}, \citenamefont {Dubovsky}, \citenamefont {Kaloper},\ and\ \citenamefont {March-Russell}}]{majorons1}%
  \BibitemOpen
  \bibfield  {author} {\bibinfo {author} {\bibfnamefont {A.}~\bibnamefont {Arvanitaki}}, \bibinfo {author} {\bibfnamefont {S.}~\bibnamefont {Dimopoulos}}, \bibinfo {author} {\bibfnamefont {S.}~\bibnamefont {Dubovsky}}, \bibinfo {author} {\bibfnamefont {N.}~\bibnamefont {Kaloper}},\ and\ \bibinfo {author} {\bibfnamefont {J.}~\bibnamefont {March-Russell}},\ }\bibfield  {title} {\bibinfo {title} {String axiverse},\ }\href {https://doi.org/10.1103/PhysRevD.81.123530} {\bibfield  {journal} {\bibinfo  {journal} {Phys. Rev. D}\ }\textbf {\bibinfo {volume} {81}},\ \bibinfo {pages} {123530} (\bibinfo {year} {2010})}\BibitemShut {NoStop}%
\bibitem [{\citenamefont {Appelquist}\ \emph {et~al.}(2003)\citenamefont {Appelquist}, \citenamefont {Dobrescu},\ and\ \citenamefont {Hopper}}]{zboson1}%
  \BibitemOpen
  \bibfield  {author} {\bibinfo {author} {\bibfnamefont {T.}~\bibnamefont {Appelquist}}, \bibinfo {author} {\bibfnamefont {B.~A.}\ \bibnamefont {Dobrescu}},\ and\ \bibinfo {author} {\bibfnamefont {A.~R.}\ \bibnamefont {Hopper}},\ }\bibfield  {title} {\bibinfo {title} {Nonexotic neutral gauge bosons},\ }\href {https://doi.org/10.1103/PhysRevD.68.035012} {\bibfield  {journal} {\bibinfo  {journal} {Phys. Rev. D}\ }\textbf {\bibinfo {volume} {68}},\ \bibinfo {pages} {035012} (\bibinfo {year} {2003})}\BibitemShut {NoStop}%
\bibitem [{\citenamefont {Langacker}(2009)}]{zboson2}%
  \BibitemOpen
  \bibfield  {author} {\bibinfo {author} {\bibfnamefont {P.}~\bibnamefont {Langacker}},\ }\bibfield  {title} {\bibinfo {title} {The physics of heavy z' gauge bosons},\ }\href {https://doi.org/10.1103/RevModPhys.81.1199} {\bibfield  {journal} {\bibinfo  {journal} {Rev. Mod. Phys.}\ }\textbf {\bibinfo {volume} {81}},\ \bibinfo {pages} {1199} (\bibinfo {year} {2009})}\BibitemShut {NoStop}%
\bibitem [{\citenamefont {Fayet}(1996)}]{s1boson1}%
  \BibitemOpen
  \bibfield  {author} {\bibinfo {author} {\bibfnamefont {P.}~\bibnamefont {Fayet}},\ }\bibfield  {title} {\bibinfo {title} {New interactions and the standard models},\ }\href {https://doi.org/10.1088/0264-9391/13/11A/004} {\bibfield  {journal} {\bibinfo  {journal} {Class. Quantum Grav.}\ }\textbf {\bibinfo {volume} {13}},\ \bibinfo {pages} {A19} (\bibinfo {year} {1996})}\BibitemShut {NoStop}%
\bibitem [{\citenamefont {Fayet}(1980)}]{s1boson2}%
  \BibitemOpen
  \bibfield  {author} {\bibinfo {author} {\bibfnamefont {P.}~\bibnamefont {Fayet}},\ }\bibfield  {title} {\bibinfo {title} {Effects of the spin-1 partner of the goldstino (gravitino) on the neutral current phenomenology},\ }\href {https://doi.org/10.1016/0370-2693(80)90488-8} {\bibfield  {journal} {\bibinfo  {journal} {Phys. Lett. B}\ }\textbf {\bibinfo {volume} {95}},\ \bibinfo {pages} {285} (\bibinfo {year} {1980})}\BibitemShut {NoStop}%
\bibitem [{\citenamefont {Fayet}(1981)}]{s1boson3}%
  \BibitemOpen
  \bibfield  {author} {\bibinfo {author} {\bibfnamefont {P.}~\bibnamefont {Fayet}},\ }\bibfield  {title} {\bibinfo {title} {A la recherche d'un nouveau boson de spin un},\ }\href {https://doi.org/10.1016/0550-3213(81)90122-X} {\bibfield  {journal} {\bibinfo  {journal} {Nucl. Phys. B}\ }\textbf {\bibinfo {volume} {187}},\ \bibinfo {pages} {184} (\bibinfo {year} {1981})}\BibitemShut {NoStop}%
\bibitem [{\citenamefont {Fayet}(1989)}]{s1boson4}%
  \BibitemOpen
  \bibfield  {author} {\bibinfo {author} {\bibfnamefont {P.}~\bibnamefont {Fayet}},\ }\bibfield  {title} {\bibinfo {title} {The fifth force charge as a linear combination of baryonic, leptonic (or b-l) and the electric charges},\ }\href {https://doi.org/10.1016/0370-2693(89)91294-X} {\bibfield  {journal} {\bibinfo  {journal} {Phys. Lett. B}\ }\textbf {\bibinfo {volume} {227}},\ \bibinfo {pages} {127} (\bibinfo {year} {1989})}\BibitemShut {NoStop}%
\bibitem [{\citenamefont {Fayet}(1990)}]{s1boson5}%
  \BibitemOpen
  \bibfield  {author} {\bibinfo {author} {\bibfnamefont {P.}~\bibnamefont {Fayet}},\ }\bibfield  {title} {\bibinfo {title} {Extra u(1)'s and new forces},\ }\href {https://doi.org/10.1016/0550-3213(90)90381-M} {\bibfield  {journal} {\bibinfo  {journal} {Nucl. Phys. B}\ }\textbf {\bibinfo {volume} {347}},\ \bibinfo {pages} {743} (\bibinfo {year} {1990})}\BibitemShut {NoStop}%
\bibitem [{\citenamefont {Dobrescu}\ and\ \citenamefont {Mocioiu}(2006)}]{dobrescu_spin-dependent_2006}%
  \BibitemOpen
  \bibfield  {author} {\bibinfo {author} {\bibfnamefont {B.~A.}\ \bibnamefont {Dobrescu}}\ and\ \bibinfo {author} {\bibfnamefont {I.}~\bibnamefont {Mocioiu}},\ }\bibfield  {title} {\bibinfo {title} {Spin-dependent macroscopic forces from new particle exchange},\ }\href {https://doi.org/10.1088/1126-6708/2006/11/005} {\bibfield  {journal} {\bibinfo  {journal} {J. High Energy Phys.}\ }\textbf {\bibinfo {volume} {2006}}\bibinfo  {number} { (11)},\ \bibinfo {pages} {005}}\BibitemShut {NoStop}%
\bibitem [{\citenamefont {Fadeev}\ \emph {et~al.}(2019)\citenamefont {Fadeev}, \citenamefont {Stadnik}, \citenamefont {Ficek}, \citenamefont {Kozlov}, \citenamefont {Flambaum},\ and\ \citenamefont {Budker}}]{fadeev_revisiting_2019}%
  \BibitemOpen
\bibfield  {number} {  }\bibfield  {author} {\bibinfo {author} {\bibfnamefont {P.}~\bibnamefont {Fadeev}}, \bibinfo {author} {\bibfnamefont {Y.~V.}\ \bibnamefont {Stadnik}}, \bibinfo {author} {\bibfnamefont {F.}~\bibnamefont {Ficek}}, \bibinfo {author} {\bibfnamefont {M.~G.}\ \bibnamefont {Kozlov}}, \bibinfo {author} {\bibfnamefont {V.~V.}\ \bibnamefont {Flambaum}},\ and\ \bibinfo {author} {\bibfnamefont {D.}~\bibnamefont {Budker}},\ }\bibfield  {title} {\bibinfo {title} {Revisiting spin-dependent forces mediated by new bosons: {Potentials} in the coordinate-space representation for macroscopic- and atomic-scale experiments},\ }\href {https://doi.org/10.1103/PhysRevA.99.022113} {\bibfield  {journal} {\bibinfo  {journal} {Phys. Rev. A}\ }\textbf {\bibinfo {volume} {99}},\ \bibinfo {pages} {022113} (\bibinfo {year} {2019})}\BibitemShut {NoStop}%
\bibitem [{\citenamefont {Adelberger}\ \emph {et~al.}(2009)\citenamefont {Adelberger}, \citenamefont {Gundlach}, \citenamefont {Heckel}, \citenamefont {Hoedl},\ and\ \citenamefont {Schlamminger}}]{Adelberger_rev}%
  \BibitemOpen
  \bibfield  {author} {\bibinfo {author} {\bibfnamefont {E.}~\bibnamefont {Adelberger}}, \bibinfo {author} {\bibfnamefont {J.}~\bibnamefont {Gundlach}}, \bibinfo {author} {\bibfnamefont {B.}~\bibnamefont {Heckel}}, \bibinfo {author} {\bibfnamefont {S.}~\bibnamefont {Hoedl}},\ and\ \bibinfo {author} {\bibfnamefont {S.}~\bibnamefont {Schlamminger}},\ }\bibfield  {title} {\bibinfo {title} {Torsion balance experiments: A low-energy frontier of particle physics},\ }\href {https://doi.org/https://doi.org/10.1016/j.ppnp.2008.08.002} {\bibfield  {journal} {\bibinfo  {journal} {Prog. Part. Nucl. Phys.}\ }\textbf {\bibinfo {volume} {62}},\ \bibinfo {pages} {102} (\bibinfo {year} {2009})}\BibitemShut {NoStop}%
\bibitem [{\citenamefont {Antoniadis}\ \emph {et~al.}(2011)\citenamefont {Antoniadis}, \citenamefont {Baessler}, \citenamefont {B\"uchner}, \citenamefont {Fedorov}, \citenamefont {Hoedl}, \citenamefont {Lambrecht}, \citenamefont {Nesvizhevsky}, \citenamefont {Pignol}, \citenamefont {Protasov}, \citenamefont {Reynaud},\ and\ \citenamefont {Sobolev}}]{Antoniadis_rev}%
  \BibitemOpen
  \bibfield  {author} {\bibinfo {author} {\bibfnamefont {I.}~\bibnamefont {Antoniadis}}, \bibinfo {author} {\bibfnamefont {S.}~\bibnamefont {Baessler}}, \bibinfo {author} {\bibfnamefont {M.}~\bibnamefont {B\"uchner}}, \bibinfo {author} {\bibfnamefont {V.}~\bibnamefont {Fedorov}}, \bibinfo {author} {\bibfnamefont {S.}~\bibnamefont {Hoedl}}, \bibinfo {author} {\bibfnamefont {A.}~\bibnamefont {Lambrecht}}, \bibinfo {author} {\bibfnamefont {V.}~\bibnamefont {Nesvizhevsky}}, \bibinfo {author} {\bibfnamefont {G.}~\bibnamefont {Pignol}}, \bibinfo {author} {\bibfnamefont {K.}~\bibnamefont {Protasov}}, \bibinfo {author} {\bibfnamefont {S.}~\bibnamefont {Reynaud}},\ and\ \bibinfo {author} {\bibfnamefont {Y.}~\bibnamefont {Sobolev}},\ }\bibfield  {title} {\bibinfo {title} {Short-range fundamental forces},\ }\href {https://doi.org/10.1016/j.crhy.2011.05.004} {\bibfield  {journal} {\bibinfo  {journal} {C. R. Phys.}\ }\textbf {\bibinfo {volume} {12}},\ \bibinfo {pages} {755} (\bibinfo {year} {2011})}\BibitemShut {NoStop}%
\bibitem [{\citenamefont {Safronova}\ \emph {et~al.}(2018)\citenamefont {Safronova}, \citenamefont {Budker}, \citenamefont {DeMille}, \citenamefont {Kimball}, \citenamefont {Derevianko},\ and\ \citenamefont {Clark}}]{Safronova_rev}%
  \BibitemOpen
  \bibfield  {author} {\bibinfo {author} {\bibfnamefont {M.~S.}\ \bibnamefont {Safronova}}, \bibinfo {author} {\bibfnamefont {D.}~\bibnamefont {Budker}}, \bibinfo {author} {\bibfnamefont {D.}~\bibnamefont {DeMille}}, \bibinfo {author} {\bibfnamefont {D.~F.~J.}\ \bibnamefont {Kimball}}, \bibinfo {author} {\bibfnamefont {A.}~\bibnamefont {Derevianko}},\ and\ \bibinfo {author} {\bibfnamefont {C.~W.}\ \bibnamefont {Clark}},\ }\bibfield  {title} {\bibinfo {title} {Search for new physics with atoms and molecules},\ }\href {https://doi.org/10.1103/RevModPhys.90.025008} {\bibfield  {journal} {\bibinfo  {journal} {Rev. Mod. Phys.}\ }\textbf {\bibinfo {volume} {90}},\ \bibinfo {pages} {025008} (\bibinfo {year} {2018})}\BibitemShut {NoStop}%
\bibitem [{\citenamefont {Cong}\ \emph {et~al.}(2025)\citenamefont {Cong}, \citenamefont {Ji}, \citenamefont {Fadeev}, \citenamefont {Ficek}, \citenamefont {Jiang}, \citenamefont {Flambaum}, \citenamefont {Guan}, \citenamefont {Jackson~Kimball}, \citenamefont {Kozlov}, \citenamefont {Stadnik},\ and\ \citenamefont {Budker}}]{Cong_rev}%
  \BibitemOpen
  \bibfield  {author} {\bibinfo {author} {\bibfnamefont {L.}~\bibnamefont {Cong}}, \bibinfo {author} {\bibfnamefont {W.}~\bibnamefont {Ji}}, \bibinfo {author} {\bibfnamefont {P.}~\bibnamefont {Fadeev}}, \bibinfo {author} {\bibfnamefont {F.}~\bibnamefont {Ficek}}, \bibinfo {author} {\bibfnamefont {M.}~\bibnamefont {Jiang}}, \bibinfo {author} {\bibfnamefont {V.~V.}\ \bibnamefont {Flambaum}}, \bibinfo {author} {\bibfnamefont {H.}~\bibnamefont {Guan}}, \bibinfo {author} {\bibfnamefont {D.~F.}\ \bibnamefont {Jackson~Kimball}}, \bibinfo {author} {\bibfnamefont {M.~G.}\ \bibnamefont {Kozlov}}, \bibinfo {author} {\bibfnamefont {Y.~V.}\ \bibnamefont {Stadnik}},\ and\ \bibinfo {author} {\bibfnamefont {D.}~\bibnamefont {Budker}},\ }\bibfield  {title} {\bibinfo {title} {Spin-dependent exotic interactions},\ }\href {https://doi.org/10.1103/RevModPhys.97.025005} {\bibfield  {journal} {\bibinfo  {journal} {Rev. Mod. Phys.}\ }\textbf {\bibinfo {volume} {97}},\ \bibinfo {pages} {025005} (\bibinfo {year} {2025})}\BibitemShut
  {NoStop}%
\bibitem [{\citenamefont {Piegsa}\ and\ \citenamefont {Pignol}(2012)}]{Piegsa2012}%
  \BibitemOpen
  \bibfield  {author} {\bibinfo {author} {\bibfnamefont {F.~M.}\ \bibnamefont {Piegsa}}\ and\ \bibinfo {author} {\bibfnamefont {G.}~\bibnamefont {Pignol}},\ }\bibfield  {title} {\bibinfo {title} {Limits on the axial coupling constant of new light bosons},\ }\href {https://doi.org/10.1103/PhysRevLett.108.181801} {\bibfield  {journal} {\bibinfo  {journal} {Phys. Rev. Lett.}\ }\textbf {\bibinfo {volume} {108}},\ \bibinfo {pages} {181801} (\bibinfo {year} {2012})}\BibitemShut {NoStop}%
\bibitem [{\citenamefont {Yan}\ and\ \citenamefont {Snow}(2013)}]{Yan13}%
  \BibitemOpen
  \bibfield  {author} {\bibinfo {author} {\bibfnamefont {H.}~\bibnamefont {Yan}}\ and\ \bibinfo {author} {\bibfnamefont {W.~M.}\ \bibnamefont {Snow}},\ }\bibfield  {title} {\bibinfo {title} {New limit on possible long-range parity-odd interactions of the neutron from neutron-spin rotation in liquid $^{4}\mathrm{He}$},\ }\href {https://doi.org/10.1103/PhysRevLett.110.082003} {\bibfield  {journal} {\bibinfo  {journal} {Phys. Rev. Lett.}\ }\textbf {\bibinfo {volume} {110}},\ \bibinfo {pages} {082003} (\bibinfo {year} {2013})}\BibitemShut {NoStop}%
\bibitem [{\citenamefont {Lehnert}\ \emph {et~al.}(2017)\citenamefont {Lehnert}, \citenamefont {Snow}, \citenamefont {Xiao},\ and\ \citenamefont {Xu}}]{Lehnert17}%
  \BibitemOpen
  \bibfield  {author} {\bibinfo {author} {\bibfnamefont {R.}~\bibnamefont {Lehnert}}, \bibinfo {author} {\bibfnamefont {W.~M.}\ \bibnamefont {Snow}}, \bibinfo {author} {\bibfnamefont {Z.}~\bibnamefont {Xiao}},\ and\ \bibinfo {author} {\bibfnamefont {R.}~\bibnamefont {Xu}},\ }\bibfield  {title} {\bibinfo {title} {Constraining spacetime nonmetricity with neutron spin rotation in liquid {4He}},\ }\href {https://doi.org/https://doi.org/10.1016/j.physletb.2017.07.059} {\bibfield  {journal} {\bibinfo  {journal} {Phys. Lett. B}\ }\textbf {\bibinfo {volume} {772}},\ \bibinfo {pages} {865} (\bibinfo {year} {2017})}\BibitemShut {NoStop}%
\bibitem [{\citenamefont {Parnell}\ \emph {et~al.}(2020)\citenamefont {Parnell}, \citenamefont {van Well}, \citenamefont {Plomp}, \citenamefont {Dalgliesh}, \citenamefont {Juliane-Steinke}, \citenamefont {Cooper}, \citenamefont {Geerits}, \citenamefont {de~Haan}, \citenamefont {Steffen},\ and\ \citenamefont {Snow}}]{Parnell:2020}%
  \BibitemOpen
  \bibfield  {author} {\bibinfo {author} {\bibfnamefont {S.~R.}\ \bibnamefont {Parnell}}, \bibinfo {author} {\bibfnamefont {A.}~\bibnamefont {van Well}}, \bibinfo {author} {\bibfnamefont {J.}~\bibnamefont {Plomp}}, \bibinfo {author} {\bibfnamefont {R.~M.}\ \bibnamefont {Dalgliesh}}, \bibinfo {author} {\bibfnamefont {N.}~\bibnamefont {Juliane-Steinke}}, \bibinfo {author} {\bibfnamefont {J.~K.}\ \bibnamefont {Cooper}}, \bibinfo {author} {\bibfnamefont {N.}~\bibnamefont {Geerits}}, \bibinfo {author} {\bibfnamefont {V.~O.}\ \bibnamefont {de~Haan}}, \bibinfo {author} {\bibfnamefont {K.~E.}\ \bibnamefont {Steffen}},\ and\ \bibinfo {author} {\bibfnamefont {W.~M.}\ \bibnamefont {Snow}},\ }\bibfield  {title} {\bibinfo {title} {{A Search for exotic spin-dependent couplings of the neutron with matter using spin-echo based neutron interferometry}},\ }\href@noop {} {\bibfield  {journal} {\bibinfo  {journal} {Phys. Rev. D}\ }\textbf {\bibinfo {volume} {101}},\ \bibinfo {pages} {122002} (\bibinfo {year} {2020})}\BibitemShut
  {NoStop}%
\bibitem [{\citenamefont {Mulkey}\ \emph {et~al.}(2026)\citenamefont {Mulkey} \emph {et~al.}}]{Mulkey:2025wnb}%
  \BibitemOpen
  \bibfield  {author} {\bibinfo {author} {\bibfnamefont {T.}~\bibnamefont {Mulkey}} \emph {et~al.},\ }\bibfield  {title} {\bibinfo {title} {{New Upper Bounds on Exotic Neutron-Spin{\textendash}Electron-Spin Interactions via Neutron-Spin-Rotation Measurements in a Compensated Ferrimagnet}},\ }\href {https://doi.org/10.1103/bjw2-fbfc} {\bibfield  {journal} {\bibinfo  {journal} {Phys. Rev. Lett.}\ }\textbf {\bibinfo {volume} {136}},\ \bibinfo {pages} {071801} (\bibinfo {year} {2026})},\ \Eprint {https://arxiv.org/abs/2508.14243} {arXiv:2508.14243 [nucl-ex]} \BibitemShut {NoStop}%
\bibitem [{\citenamefont {Michel}(1964)}]{Michel:1964}%
  \BibitemOpen
  \bibfield  {author} {\bibinfo {author} {\bibfnamefont {F.~C.}\ \bibnamefont {Michel}},\ }\href@noop {} {\bibfield  {journal} {\bibinfo  {journal} {Physical Review}\ }\textbf {\bibinfo {volume} {133}},\ \bibinfo {pages} {B329} (\bibinfo {year} {1964})}\BibitemShut {NoStop}%
\bibitem [{\citenamefont {Stodolsky}(1982)}]{Stodolsky}%
  \BibitemOpen
  \bibfield  {author} {\bibinfo {author} {\bibfnamefont {L.}~\bibnamefont {Stodolsky}},\ }\href@noop {} {\bibfield  {journal} {\bibinfo  {journal} {Nuclear Physics B}\ }\textbf {\bibinfo {volume} {197}},\ \bibinfo {pages} {213} (\bibinfo {year} {1982})}\BibitemShut {NoStop}%
\bibitem [{\citenamefont {Sushkov}\ and\ \citenamefont {Flambaum}(1980)}]{Sushkov1980}%
  \BibitemOpen
  \bibfield  {author} {\bibinfo {author} {\bibfnamefont {O.}~\bibnamefont {Sushkov}}\ and\ \bibinfo {author} {\bibfnamefont {V.}~\bibnamefont {Flambaum}},\ }\bibfield  {title} {\bibinfo {title} {Parity violation in neutron capture reactions},\ }\href@noop {} {\bibfield  {journal} {\bibinfo  {journal} {JETP Letters}\ }\textbf {\bibinfo {volume} {32}},\ \bibinfo {pages} {353} (\bibinfo {year} {1980})}\BibitemShut {NoStop}%
\bibitem [{\citenamefont {Sushkov}\ and\ \citenamefont {Flambaum}(1982)}]{Sushkov1982}%
  \BibitemOpen
  \bibfield  {author} {\bibinfo {author} {\bibfnamefont {O.}~\bibnamefont {Sushkov}}\ and\ \bibinfo {author} {\bibfnamefont {V.}~\bibnamefont {Flambaum}},\ }\bibfield  {title} {\bibinfo {title} {Nuclear parity nonconservation},\ }\href@noop {} {\bibfield  {journal} {\bibinfo  {journal} {Sov. Phys. Usp.}\ }\textbf {\bibinfo {volume} {25}},\ \bibinfo {pages} {1} (\bibinfo {year} {1982})}\BibitemShut {NoStop}%
\bibitem [{\citenamefont {Forte}\ \emph {et~al.}(1980)\citenamefont {Forte}, \citenamefont {Heckel}, \citenamefont {Ramsey}, \citenamefont {Green}, \citenamefont {Greene}, \citenamefont {Byrne},\ and\ \citenamefont {Pendlebury}}]{Forte80}%
  \BibitemOpen
  \bibfield  {author} {\bibinfo {author} {\bibfnamefont {M.}~\bibnamefont {Forte}}, \bibinfo {author} {\bibfnamefont {B.~R.}\ \bibnamefont {Heckel}}, \bibinfo {author} {\bibfnamefont {N.~F.}\ \bibnamefont {Ramsey}}, \bibinfo {author} {\bibfnamefont {K.}~\bibnamefont {Green}}, \bibinfo {author} {\bibfnamefont {G.~L.}\ \bibnamefont {Greene}}, \bibinfo {author} {\bibfnamefont {J.}~\bibnamefont {Byrne}},\ and\ \bibinfo {author} {\bibfnamefont {J.~M.}\ \bibnamefont {Pendlebury}},\ }\bibfield  {title} {\bibinfo {title} {First measurement of parity-nonconserving neutron-spin rotation: The tin isotope},\ }\href@noop {} {\bibfield  {journal} {\bibinfo  {journal} {Phys. Rev. Lett.}\ }\textbf {\bibinfo {volume} {45}},\ \bibinfo {pages} {2088} (\bibinfo {year} {1980})}\BibitemShut {NoStop}%
\bibitem [{\citenamefont {Alfimenkov}\ \emph {et~al.}(1983)\citenamefont {Alfimenkov}, \citenamefont {Borzakov}, \citenamefont {{Van Thuan}}, \citenamefont {Mareev}, \citenamefont {Pikelner}, \citenamefont {Khrykin},\ and\ \citenamefont {Sharapov}}]{Alfimenkov1983}%
  \BibitemOpen
  \bibfield  {author} {\bibinfo {author} {\bibfnamefont {V.}~\bibnamefont {Alfimenkov}}, \bibinfo {author} {\bibfnamefont {S.}~\bibnamefont {Borzakov}}, \bibinfo {author} {\bibfnamefont {V.}~\bibnamefont {{Van Thuan}}}, \bibinfo {author} {\bibfnamefont {Y.}~\bibnamefont {Mareev}}, \bibinfo {author} {\bibfnamefont {L.}~\bibnamefont {Pikelner}}, \bibinfo {author} {\bibfnamefont {A.}~\bibnamefont {Khrykin}},\ and\ \bibinfo {author} {\bibfnamefont {E.}~\bibnamefont {Sharapov}},\ }\bibfield  {title} {\bibinfo {title} {Parity nonconservation in neutron resonances},\ }\href {https://doi.org/https://doi.org/10.1016/0375-9474(83)90649-8} {\bibfield  {journal} {\bibinfo  {journal} {Nucl. Phys. A}\ }\textbf {\bibinfo {volume} {398}},\ \bibinfo {pages} {93} (\bibinfo {year} {1983})}\BibitemShut {NoStop}%
\bibitem [{\citenamefont {Alfimenkov}\ \emph {et~al.}(1984)\citenamefont {Alfimenkov}, \citenamefont {Borzakov}, \citenamefont {Van~Tkhuan}, \citenamefont {Mareev}, \citenamefont {Pikelner}, \citenamefont {Frank}, \citenamefont {Khrykin},\ and\ \citenamefont {Sharapov}}]{Alfimenkov1984}%
  \BibitemOpen
  \bibfield  {author} {\bibinfo {author} {\bibfnamefont {V.~P.}\ \bibnamefont {Alfimenkov}}, \bibinfo {author} {\bibfnamefont {S.~B.}\ \bibnamefont {Borzakov}}, \bibinfo {author} {\bibfnamefont {V.}~\bibnamefont {Van~Tkhuan}}, \bibinfo {author} {\bibfnamefont {Y.~D.}\ \bibnamefont {Mareev}}, \bibinfo {author} {\bibfnamefont {L.~B.}\ \bibnamefont {Pikelner}}, \bibinfo {author} {\bibfnamefont {I.~M.}\ \bibnamefont {Frank}}, \bibinfo {author} {\bibfnamefont {A.~S.}\ \bibnamefont {Khrykin}},\ and\ \bibinfo {author} {\bibfnamefont {E.~I.}\ \bibnamefont {Sharapov}},\ }\href@noop {} {\bibfield  {journal} {\bibinfo  {journal} {JETP Lett.}\ }\textbf {\bibinfo {volume} {39}},\ \bibinfo {pages} {416} (\bibinfo {year} {1984})}\BibitemShut {NoStop}%
\bibitem [{\citenamefont {Heckel}\ \emph {et~al.}(1982)\citenamefont {Heckel} \emph {et~al.}}]{Heckel1982}%
  \BibitemOpen
  \bibfield  {author} {\bibinfo {author} {\bibfnamefont {B.}~\bibnamefont {Heckel}} \emph {et~al.},\ }\bibfield  {title} {\bibinfo {title} {A measurement of parity nonconserving neutron spin rotation in lead and tin},\ }\href@noop {} {\bibfield  {journal} {\bibinfo  {journal} {Physics Letters B}\ }\textbf {\bibinfo {volume} {119}},\ \bibinfo {pages} {298} (\bibinfo {year} {1982})}\BibitemShut {NoStop}%
\bibitem [{\citenamefont {Heckel}\ \emph {et~al.}(1984)\citenamefont {Heckel} \emph {et~al.}}]{Heckel1984}%
  \BibitemOpen
  \bibfield  {author} {\bibinfo {author} {\bibfnamefont {B.}~\bibnamefont {Heckel}} \emph {et~al.},\ }\bibfield  {title} {\bibinfo {title} {Measurement of parity nonconserving neutron spin rotation in lanthanum},\ }\href@noop {} {\bibfield  {journal} {\bibinfo  {journal} {Physical Review C}\ }\textbf {\bibinfo {volume} {29}},\ \bibinfo {pages} {2389} (\bibinfo {year} {1984})}\BibitemShut {NoStop}%
\bibitem [{\citenamefont {Yuan}\ \emph {et~al.}(1991)\citenamefont {Yuan}, \citenamefont {Bowman}, \citenamefont {Bowman}, \citenamefont {Bush}, \citenamefont {Delheij}, \citenamefont {Frankle}, \citenamefont {Gould}, \citenamefont {Haase}, \citenamefont {Knudson}, \citenamefont {Mitchell}, \citenamefont {Penttil\"a}, \citenamefont {Postma}, \citenamefont {Roberson}, \citenamefont {Seestrom}, \citenamefont {Szymanski},\ and\ \citenamefont {Zhu}}]{Yuan1991}%
  \BibitemOpen
  \bibfield  {author} {\bibinfo {author} {\bibfnamefont {V.~W.}\ \bibnamefont {Yuan}}, \bibinfo {author} {\bibfnamefont {C.~D.}\ \bibnamefont {Bowman}}, \bibinfo {author} {\bibfnamefont {J.~D.}\ \bibnamefont {Bowman}}, \bibinfo {author} {\bibfnamefont {J.~E.}\ \bibnamefont {Bush}}, \bibinfo {author} {\bibfnamefont {P.~P.~J.}\ \bibnamefont {Delheij}}, \bibinfo {author} {\bibfnamefont {C.~M.}\ \bibnamefont {Frankle}}, \bibinfo {author} {\bibfnamefont {C.~R.}\ \bibnamefont {Gould}}, \bibinfo {author} {\bibfnamefont {D.~G.}\ \bibnamefont {Haase}}, \bibinfo {author} {\bibfnamefont {J.~N.}\ \bibnamefont {Knudson}}, \bibinfo {author} {\bibfnamefont {G.~E.}\ \bibnamefont {Mitchell}}, \bibinfo {author} {\bibfnamefont {S.}~\bibnamefont {Penttil\"a}}, \bibinfo {author} {\bibfnamefont {H.}~\bibnamefont {Postma}}, \bibinfo {author} {\bibfnamefont {N.~R.}\ \bibnamefont {Roberson}}, \bibinfo {author} {\bibfnamefont {S.~J.}\ \bibnamefont {Seestrom}}, \bibinfo {author} {\bibfnamefont {J.~J.}\ \bibnamefont {Szymanski}},\ and\
  \bibinfo {author} {\bibfnamefont {X.}~\bibnamefont {Zhu}},\ }\bibfield  {title} {\bibinfo {title} {Parity nonconservation in polarized-neutron transmission through $^{139}\mathrm{La}$},\ }\href {https://doi.org/10.1103/PhysRevC.44.2187} {\bibfield  {journal} {\bibinfo  {journal} {Phys. Rev. C}\ }\textbf {\bibinfo {volume} {44}},\ \bibinfo {pages} {2187} (\bibinfo {year} {1991})}\BibitemShut {NoStop}%
\bibitem [{\citenamefont {Heil}\ \emph {et~al.}(1999)\citenamefont {Heil}, \citenamefont {Humblot}, \citenamefont {Hofmann}, \citenamefont {Krasnoschekova}, \citenamefont {Lelievre-Berna}, \citenamefont {Petoukhov}, \citenamefont {Petrov}, \citenamefont {Serebrov}, \citenamefont {Tasset},\ and\ \citenamefont {Valsky}}]{Heil1999}%
  \BibitemOpen
  \bibfield  {author} {\bibinfo {author} {\bibfnamefont {W.}~\bibnamefont {Heil}}, \bibinfo {author} {\bibfnamefont {H.}~\bibnamefont {Humblot}}, \bibinfo {author} {\bibfnamefont {D.}~\bibnamefont {Hofmann}}, \bibinfo {author} {\bibfnamefont {I.}~\bibnamefont {Krasnoschekova}}, \bibinfo {author} {\bibfnamefont {E.}~\bibnamefont {Lelievre-Berna}}, \bibinfo {author} {\bibfnamefont {A.}~\bibnamefont {Petoukhov}}, \bibinfo {author} {\bibfnamefont {G.}~\bibnamefont {Petrov}}, \bibinfo {author} {\bibfnamefont {A.}~\bibnamefont {Serebrov}}, \bibinfo {author} {\bibfnamefont {F.}~\bibnamefont {Tasset}},\ and\ \bibinfo {author} {\bibfnamefont {G.}~\bibnamefont {Valsky}},\ }\bibfield  {title} {\bibinfo {title} {Neutron optics p-violation effects near p-wave resonance},\ }\href@noop {} {\bibfield  {journal} {\bibinfo  {journal} {Physica B}\ }\textbf {\bibinfo {volume} {267-268}},\ \bibinfo {pages} {289} (\bibinfo {year} {1999})}\BibitemShut {NoStop}%
\bibitem [{\citenamefont {Haseyama}\ \emph {et~al.}(2002)\citenamefont {Haseyama}, \citenamefont {Asahi}, \citenamefont {Bowman}, \citenamefont {Delheij}, \citenamefont {Funahashi}, \citenamefont {Ishimoto}, \citenamefont {Jones}, \citenamefont {Masaike}, \citenamefont {Masuda}, \citenamefont {Matsuda}, \citenamefont {Morimoto}, \citenamefont {Muto}, \citenamefont {Penttilä}, \citenamefont {Pomeroy}, \citenamefont {Sakai}, \citenamefont {Sharapov}, \citenamefont {Smith},\ and\ \citenamefont {Yuan}}]{Haseyama2002}%
  \BibitemOpen
  \bibfield  {author} {\bibinfo {author} {\bibfnamefont {T.}~\bibnamefont {Haseyama}}, \bibinfo {author} {\bibfnamefont {K.}~\bibnamefont {Asahi}}, \bibinfo {author} {\bibfnamefont {J.}~\bibnamefont {Bowman}}, \bibinfo {author} {\bibfnamefont {P.}~\bibnamefont {Delheij}}, \bibinfo {author} {\bibfnamefont {H.}~\bibnamefont {Funahashi}}, \bibinfo {author} {\bibfnamefont {S.}~\bibnamefont {Ishimoto}}, \bibinfo {author} {\bibfnamefont {G.}~\bibnamefont {Jones}}, \bibinfo {author} {\bibfnamefont {A.}~\bibnamefont {Masaike}}, \bibinfo {author} {\bibfnamefont {Y.}~\bibnamefont {Masuda}}, \bibinfo {author} {\bibfnamefont {Y.}~\bibnamefont {Matsuda}}, \bibinfo {author} {\bibfnamefont {K.}~\bibnamefont {Morimoto}}, \bibinfo {author} {\bibfnamefont {S.}~\bibnamefont {Muto}}, \bibinfo {author} {\bibfnamefont {S.}~\bibnamefont {Penttilä}}, \bibinfo {author} {\bibfnamefont {V.}~\bibnamefont {Pomeroy}}, \bibinfo {author} {\bibfnamefont {K.}~\bibnamefont {Sakai}}, \bibinfo {author} {\bibfnamefont {E.}~\bibnamefont
  {Sharapov}}, \bibinfo {author} {\bibfnamefont {D.}~\bibnamefont {Smith}},\ and\ \bibinfo {author} {\bibfnamefont {V.}~\bibnamefont {Yuan}},\ }\bibfield  {title} {\bibinfo {title} {Measurement of parity-nonconserving rotation of neutron spin in the 0.734-ev p-wave resonance of $^{139}$la},\ }\href {https://doi.org/https://doi.org/10.1016/S0370-2693(02)01720-3} {\bibfield  {journal} {\bibinfo  {journal} {Physics Letters B}\ }\textbf {\bibinfo {volume} {534}},\ \bibinfo {pages} {39} (\bibinfo {year} {2002})}\BibitemShut {NoStop}%
\bibitem [{\citenamefont {Mitchell}\ \emph {et~al.}(1999)\citenamefont {Mitchell}, \citenamefont {Bowman},\ and\ \citenamefont {Weidenmüller}}]{Mitchell:1999}%
  \BibitemOpen
  \bibfield  {author} {\bibinfo {author} {\bibfnamefont {G.~E.}\ \bibnamefont {Mitchell}}, \bibinfo {author} {\bibfnamefont {J.~D.}\ \bibnamefont {Bowman}},\ and\ \bibinfo {author} {\bibfnamefont {H.~A.}\ \bibnamefont {Weidenmüller}},\ }\bibfield  {title} {\bibinfo {title} {Parity violation in compound nuclear resonances},\ }\href@noop {} {\bibfield  {journal} {\bibinfo  {journal} {Reviews of Modern Physics}\ }\textbf {\bibinfo {volume} {71}},\ \bibinfo {pages} {445} (\bibinfo {year} {1999})}\BibitemShut {NoStop}%
\bibitem [{\citenamefont {Mitchell}\ \emph {et~al.}(2001)\citenamefont {Mitchell}, \citenamefont {Bowman}, \citenamefont {Penttilä},\ and\ \citenamefont {Sharapov}}]{Mitchell:2001}%
  \BibitemOpen
  \bibfield  {author} {\bibinfo {author} {\bibfnamefont {G.}~\bibnamefont {Mitchell}}, \bibinfo {author} {\bibfnamefont {D.}~\bibnamefont {Bowman}}, \bibinfo {author} {\bibfnamefont {S.}~\bibnamefont {Penttilä}},\ and\ \bibinfo {author} {\bibfnamefont {E.}~\bibnamefont {Sharapov}},\ }\bibfield  {title} {\bibinfo {title} {Parity violation in compound nuclei: Experimental methods and recent results},\ }\href {https://doi.org/10.1016/S0370-1573(01)00016-3} {\bibfield  {journal} {\bibinfo  {journal} {Physics Reports}\ }\textbf {\bibinfo {volume} {354}},\ \bibinfo {pages} {157} (\bibinfo {year} {2001})}\BibitemShut {NoStop}%
\bibitem [{\citenamefont {Swanson}\ \emph {et~al.}(2019)\citenamefont {Swanson}, \citenamefont {Nico}, \citenamefont {Snow}, \citenamefont {Bass}, \citenamefont {Bass}, \citenamefont {Crawford}, \citenamefont {Gan}, \citenamefont {Heckel}, \citenamefont {Luo}, \citenamefont {Markoff}, \citenamefont {Micherdzinska}, \citenamefont {Mumm}, \citenamefont {Sarsour}, \citenamefont {Sharapov}, \citenamefont {Walbridge},\ and\ \citenamefont {Zhumabekova}}]{Swanson:2019}%
  \BibitemOpen
  \bibfield  {author} {\bibinfo {author} {\bibfnamefont {H.~E.}\ \bibnamefont {Swanson}}, \bibinfo {author} {\bibfnamefont {J.~S.}\ \bibnamefont {Nico}}, \bibinfo {author} {\bibfnamefont {W.~M.}\ \bibnamefont {Snow}}, \bibinfo {author} {\bibfnamefont {C.~D.}\ \bibnamefont {Bass}}, \bibinfo {author} {\bibfnamefont {T.~D.}\ \bibnamefont {Bass}}, \bibinfo {author} {\bibfnamefont {B.~E.}\ \bibnamefont {Crawford}}, \bibinfo {author} {\bibfnamefont {K.~F.}\ \bibnamefont {Gan}}, \bibinfo {author} {\bibfnamefont {B.~R.}\ \bibnamefont {Heckel}}, \bibinfo {author} {\bibfnamefont {D.}~\bibnamefont {Luo}}, \bibinfo {author} {\bibfnamefont {D.~M.}\ \bibnamefont {Markoff}}, \bibinfo {author} {\bibfnamefont {A.~M.}\ \bibnamefont {Micherdzinska}}, \bibinfo {author} {\bibfnamefont {H.~P.}\ \bibnamefont {Mumm}}, \bibinfo {author} {\bibfnamefont {M.}~\bibnamefont {Sarsour}}, \bibinfo {author} {\bibfnamefont {E.~I.}\ \bibnamefont {Sharapov}}, \bibinfo {author} {\bibfnamefont {S.~B.}\ \bibnamefont {Walbridge}},\ and\ \bibinfo
  {author} {\bibfnamefont {V.}~\bibnamefont {Zhumabekova}},\ }\bibfield  {title} {\bibinfo {title} {{An Upper Bound on Parity Violating Neutron Spin Rotation in $^{4}$He}},\ }\href@noop {} {\bibfield  {journal} {\bibinfo  {journal} {Phys. Rev. C}\ }\textbf {\bibinfo {volume} {100}},\ \bibinfo {pages} {015204} (\bibinfo {year} {2019})},\ \Eprint {https://arxiv.org/abs/1902.08026} {arXiv:1902.08026} \BibitemShut {NoStop}%
\bibitem [{\citenamefont {Bowman}\ \emph {et~al.}(1993)\citenamefont {Bowman} \emph {et~al.}}]{Bowman:1993}%
  \BibitemOpen
  \bibfield  {author} {\bibinfo {author} {\bibfnamefont {J.~D.}\ \bibnamefont {Bowman}} \emph {et~al.},\ }\bibfield  {title} {\bibinfo {title} {Experimental limit on parity violation in nonresonant neutron-nucleus scattering},\ }\href@noop {} {\bibfield  {journal} {\bibinfo  {journal} {Physical Review C}\ }\textbf {\bibinfo {volume} {48}},\ \bibinfo {pages} {1116} (\bibinfo {year} {1993})}\BibitemShut {NoStop}%
\bibitem [{\citenamefont {Mitchell}\ \emph {et~al.}(2000)\citenamefont {Mitchell}, \citenamefont {Bowman}, \citenamefont {Penttil\"a},\ and\ \citenamefont {Sharapov}}]{Mitchell:2000}%
  \BibitemOpen
  \bibfield  {author} {\bibinfo {author} {\bibfnamefont {G.~E.}\ \bibnamefont {Mitchell}}, \bibinfo {author} {\bibfnamefont {J.~D.}\ \bibnamefont {Bowman}}, \bibinfo {author} {\bibfnamefont {S.~I.}\ \bibnamefont {Penttil\"a}},\ and\ \bibinfo {author} {\bibfnamefont {E.~I.}\ \bibnamefont {Sharapov}},\ }\bibfield  {title} {\bibinfo {title} {New search for parity violation in nonresonant neutron scattering on thorium},\ }\href@noop {} {\bibfield  {journal} {\bibinfo  {journal} {Physical Review C}\ }\textbf {\bibinfo {volume} {61}},\ \bibinfo {pages} {045503} (\bibinfo {year} {2000})}\BibitemShut {NoStop}%
\bibitem [{\citenamefont {Frankle}\ \emph {et~al.}(1992)\citenamefont {Frankle} \emph {et~al.}}]{TRIPLE:1992jrn}%
  \BibitemOpen
  \bibfield  {author} {\bibinfo {author} {\bibfnamefont {C.~M.}\ \bibnamefont {Frankle}} \emph {et~al.} (\bibinfo {collaboration} {TRIPLE}),\ }\bibfield  {title} {\bibinfo {title} {{Parity nonconservation for neutron resonances in Th-232}},\ }\href {https://doi.org/10.1103/PhysRevC.46.778} {\bibfield  {journal} {\bibinfo  {journal} {Phys. Rev. C}\ }\textbf {\bibinfo {volume} {46}},\ \bibinfo {pages} {778} (\bibinfo {year} {1992})}\BibitemShut {NoStop}%
\bibitem [{\citenamefont {Stephenson}\ \emph {et~al.}(1998)\citenamefont {Stephenson}, \citenamefont {Bowman}, \citenamefont {Crawford}, \citenamefont {Delheij}, \citenamefont {Frankle}, \citenamefont {Iinuma}, \citenamefont {Knudson}, \citenamefont {Lowie}, \citenamefont {Masaike}, \citenamefont {Matsuda}, \citenamefont {Mitchell}, \citenamefont {Penttil{\"a}}, \citenamefont {Postma}, \citenamefont {Roberson}, \citenamefont {Seestrom}, \citenamefont {Sharapov}, \citenamefont {Yen},\ and\ \citenamefont {Yuan}}]{PhysRevC.58.1236}%
  \BibitemOpen
  \bibfield  {author} {\bibinfo {author} {\bibfnamefont {S.~L.}\ \bibnamefont {Stephenson}}, \bibinfo {author} {\bibfnamefont {J.~D.}\ \bibnamefont {Bowman}}, \bibinfo {author} {\bibfnamefont {B.~E.}\ \bibnamefont {Crawford}}, \bibinfo {author} {\bibfnamefont {P.~P.~J.}\ \bibnamefont {Delheij}}, \bibinfo {author} {\bibfnamefont {C.~M.}\ \bibnamefont {Frankle}}, \bibinfo {author} {\bibfnamefont {M.}~\bibnamefont {Iinuma}}, \bibinfo {author} {\bibfnamefont {J.~N.}\ \bibnamefont {Knudson}}, \bibinfo {author} {\bibfnamefont {L.~Y.}\ \bibnamefont {Lowie}}, \bibinfo {author} {\bibfnamefont {A.}~\bibnamefont {Masaike}}, \bibinfo {author} {\bibfnamefont {Y.}~\bibnamefont {Matsuda}}, \bibinfo {author} {\bibfnamefont {G.~E.}\ \bibnamefont {Mitchell}}, \bibinfo {author} {\bibfnamefont {S.~I.}\ \bibnamefont {Penttil{\"a}}}, \bibinfo {author} {\bibfnamefont {H.}~\bibnamefont {Postma}}, \bibinfo {author} {\bibfnamefont {N.~R.}\ \bibnamefont {Roberson}}, \bibinfo {author} {\bibfnamefont {S.~J.}\ \bibnamefont {Seestrom}},
  \bibinfo {author} {\bibfnamefont {E.~I.}\ \bibnamefont {Sharapov}}, \bibinfo {author} {\bibfnamefont {Y.-F.}\ \bibnamefont {Yen}},\ and\ \bibinfo {author} {\bibfnamefont {V.~W.}\ \bibnamefont {Yuan}},\ }\bibfield  {title} {\bibinfo {title} {Parity nonconservation in neutron resonances in ${}^{232}\mathrm{Th}$},\ }\href {https://doi.org/10.1103/PhysRevC.58.1236} {\bibfield  {journal} {\bibinfo  {journal} {Phys. Rev. C}\ }\textbf {\bibinfo {volume} {58}},\ \bibinfo {pages} {1236} (\bibinfo {year} {1998})}\BibitemShut {NoStop}%
\bibitem [{\citenamefont {Blyth}\ \emph {et~al.}(2018)\citenamefont {Blyth} \emph {et~al.}}]{NPDGamma:2018vhh}%
  \BibitemOpen
  \bibfield  {author} {\bibinfo {author} {\bibfnamefont {D.}~\bibnamefont {Blyth}} \emph {et~al.} (\bibinfo {collaboration} {NPDGamma}),\ }\bibfield  {title} {\bibinfo {title} {{First Observation of $P$-odd $\gamma$ Asymmetry in Polarized Neutron Capture on Hydrogen}},\ }\href {https://doi.org/10.1103/PhysRevLett.121.242002} {\bibfield  {journal} {\bibinfo  {journal} {Phys. Rev. Lett.}\ }\textbf {\bibinfo {volume} {121}},\ \bibinfo {pages} {242002} (\bibinfo {year} {2018})},\ \Eprint {https://arxiv.org/abs/1807.10192} {arXiv:1807.10192 [nucl-ex]} \BibitemShut {NoStop}%
\bibitem [{\citenamefont {Gericke}\ \emph {et~al.}(2020)\citenamefont {Gericke}, \citenamefont {Bae\ss{}ler}, \citenamefont {Barr\'on-Palos}, \citenamefont {Birge}, \citenamefont {Bowman}, \citenamefont {Calarco}, \citenamefont {Cianciolo}, \citenamefont {Coppola}, \citenamefont {Crawford}, \citenamefont {Fomin}, \citenamefont {Garishvili}, \citenamefont {Greene}, \citenamefont {Hale}, \citenamefont {Hamblen}, \citenamefont {Hayes}, \citenamefont {Iverson}, \citenamefont {Kabir}, \citenamefont {McCrea}, \citenamefont {Plemons}, \citenamefont {Ram\'{\i}rez-Morales}, \citenamefont {Mueller}, \citenamefont {Novikov}, \citenamefont {Penttila}, \citenamefont {Scott}, \citenamefont {Watts},\ and\ \citenamefont {Wickersham}}]{Gericke2020}%
  \BibitemOpen
  \bibfield  {author} {\bibinfo {author} {\bibfnamefont {M.~T.}\ \bibnamefont {Gericke}}, \bibinfo {author} {\bibfnamefont {S.}~\bibnamefont {Bae\ss{}ler}}, \bibinfo {author} {\bibfnamefont {L.}~\bibnamefont {Barr\'on-Palos}}, \bibinfo {author} {\bibfnamefont {N.}~\bibnamefont {Birge}}, \bibinfo {author} {\bibfnamefont {J.~D.}\ \bibnamefont {Bowman}}, \bibinfo {author} {\bibfnamefont {J.}~\bibnamefont {Calarco}}, \bibinfo {author} {\bibfnamefont {V.}~\bibnamefont {Cianciolo}}, \bibinfo {author} {\bibfnamefont {C.~E.}\ \bibnamefont {Coppola}}, \bibinfo {author} {\bibfnamefont {C.~B.}\ \bibnamefont {Crawford}}, \bibinfo {author} {\bibfnamefont {N.}~\bibnamefont {Fomin}}, \bibinfo {author} {\bibfnamefont {I.}~\bibnamefont {Garishvili}}, \bibinfo {author} {\bibfnamefont {G.~L.}\ \bibnamefont {Greene}}, \bibinfo {author} {\bibfnamefont {G.~M.}\ \bibnamefont {Hale}}, \bibinfo {author} {\bibfnamefont {J.}~\bibnamefont {Hamblen}}, \bibinfo {author} {\bibfnamefont {C.}~\bibnamefont {Hayes}}, \bibinfo {author}
  {\bibfnamefont {E.}~\bibnamefont {Iverson}}, \bibinfo {author} {\bibfnamefont {M.~L.}\ \bibnamefont {Kabir}}, \bibinfo {author} {\bibfnamefont {M.}~\bibnamefont {McCrea}}, \bibinfo {author} {\bibfnamefont {E.}~\bibnamefont {Plemons}}, \bibinfo {author} {\bibfnamefont {A.}~\bibnamefont {Ram\'{\i}rez-Morales}}, \bibinfo {author} {\bibfnamefont {P.~E.}\ \bibnamefont {Mueller}}, \bibinfo {author} {\bibfnamefont {I.}~\bibnamefont {Novikov}}, \bibinfo {author} {\bibfnamefont {S.}~\bibnamefont {Penttila}}, \bibinfo {author} {\bibfnamefont {E.~M.}\ \bibnamefont {Scott}}, \bibinfo {author} {\bibfnamefont {J.}~\bibnamefont {Watts}},\ and\ \bibinfo {author} {\bibfnamefont {C.}~\bibnamefont {Wickersham}} (\bibinfo {collaboration} {$n^{3}\mathrm{He}$ Collaboration}),\ }\bibfield  {title} {\bibinfo {title} {First precision measurement of the parity violating asymmetry in cold neutron capture on $^{3}\mathrm{He}$},\ }\href {https://doi.org/10.1103/PhysRevLett.125.131803} {\bibfield  {journal} {\bibinfo  {journal} {Phys.
  Rev. Lett.}\ }\textbf {\bibinfo {volume} {125}},\ \bibinfo {pages} {131803} (\bibinfo {year} {2020})}\BibitemShut {NoStop}%
\bibitem [{\citenamefont {Desplanques}\ \emph {et~al.}(1980)\citenamefont {Desplanques}, \citenamefont {Donoghue},\ and\ \citenamefont {Holstein}}]{DDH1980}%
  \BibitemOpen
  \bibfield  {author} {\bibinfo {author} {\bibfnamefont {B.}~\bibnamefont {Desplanques}}, \bibinfo {author} {\bibfnamefont {J.~F.}\ \bibnamefont {Donoghue}},\ and\ \bibinfo {author} {\bibfnamefont {B.~R.}\ \bibnamefont {Holstein}},\ }\bibfield  {title} {\bibinfo {title} {Unified treatment of the parity violating nuclear force},\ }\href {https://doi.org/10.1016/0003-4916(80)90217-1} {\bibfield  {journal} {\bibinfo  {journal} {Annals of Physics}\ }\textbf {\bibinfo {volume} {124}},\ \bibinfo {pages} {449} (\bibinfo {year} {1980})}\BibitemShut {NoStop}%
\bibitem [{\citenamefont {Adelberger}\ and\ \citenamefont {Haxton}(1985)}]{Adelberger1985}%
  \BibitemOpen
  \bibfield  {author} {\bibinfo {author} {\bibfnamefont {E.~G.}\ \bibnamefont {Adelberger}}\ and\ \bibinfo {author} {\bibfnamefont {W.~C.}\ \bibnamefont {Haxton}},\ }\bibfield  {title} {\bibinfo {title} {Parity nonconservation in the $nn$ interaction},\ }\href {https://doi.org/10.1146/annurev.ns.35.120185.002441} {\bibfield  {journal} {\bibinfo  {journal} {Ann. Rev. Nucl. Part. Sci.}\ }\textbf {\bibinfo {volume} {35}},\ \bibinfo {pages} {501} (\bibinfo {year} {1985})}\BibitemShut {NoStop}%
\bibitem [{\citenamefont {Desplanques}(1998)}]{Desplanques:1998ak}%
  \BibitemOpen
  \bibfield  {author} {\bibinfo {author} {\bibfnamefont {B.}~\bibnamefont {Desplanques}},\ }\bibfield  {title} {\bibinfo {title} {{Parity-non-conservation in nuclear forces at low energy: Phenomenology and questions}},\ }\href {https://doi.org/10.1016/S0370-1573(97)00072-0} {\bibfield  {journal} {\bibinfo  {journal} {Phys. Rept.}\ }\textbf {\bibinfo {volume} {297}},\ \bibinfo {pages} {1} (\bibinfo {year} {1998})}\BibitemShut {NoStop}%
\bibitem [{\citenamefont {Schiavilla}\ \emph {et~al.}(2004)\citenamefont {Schiavilla}, \citenamefont {Carlson},\ and\ \citenamefont {Paris}}]{Schiavilla2004}%
  \BibitemOpen
  \bibfield  {author} {\bibinfo {author} {\bibfnamefont {R.}~\bibnamefont {Schiavilla}}, \bibinfo {author} {\bibfnamefont {J.}~\bibnamefont {Carlson}},\ and\ \bibinfo {author} {\bibfnamefont {M.}~\bibnamefont {Paris}},\ }\bibfield  {title} {\bibinfo {title} {Parity-violating interaction effects in the $np$ system},\ }\href {https://doi.org/10.1103/PhysRevC.70.044007} {\bibfield  {journal} {\bibinfo  {journal} {Phys. Rev. C}\ }\textbf {\bibinfo {volume} {70}},\ \bibinfo {pages} {044007} (\bibinfo {year} {2004})}\BibitemShut {NoStop}%
\bibitem [{\citenamefont {Zhu}\ \emph {et~al.}(2005)\citenamefont {Zhu}, \citenamefont {Maekawa}, \citenamefont {Holstein}, \citenamefont {Ramsey-Musolf},\ and\ \citenamefont {{van Kolck}}}]{Zhu2005}%
  \BibitemOpen
  \bibfield  {author} {\bibinfo {author} {\bibfnamefont {S.-L.}\ \bibnamefont {Zhu}}, \bibinfo {author} {\bibfnamefont {C.}~\bibnamefont {Maekawa}}, \bibinfo {author} {\bibfnamefont {B.}~\bibnamefont {Holstein}}, \bibinfo {author} {\bibfnamefont {M.}~\bibnamefont {Ramsey-Musolf}},\ and\ \bibinfo {author} {\bibfnamefont {U.}~\bibnamefont {{van Kolck}}},\ }\bibfield  {title} {\bibinfo {title} {Nuclear parity violation in effective field theory},\ }\href {https://doi.org/10.1016/j.nuclphysa.2004.10.032} {\bibfield  {journal} {\bibinfo  {journal} {Nucl. Phys. A}\ }\textbf {\bibinfo {volume} {748}},\ \bibinfo {pages} {435} (\bibinfo {year} {2005})}\BibitemShut {NoStop}%
\bibitem [{\citenamefont {Schiavilla}\ \emph {et~al.}(2008)\citenamefont {Schiavilla}, \citenamefont {Viviani}, \citenamefont {Girlanda}, \citenamefont {Kievsky},\ and\ \citenamefont {Marcucci}}]{Schiavilla2008}%
  \BibitemOpen
  \bibfield  {author} {\bibinfo {author} {\bibfnamefont {R.}~\bibnamefont {Schiavilla}}, \bibinfo {author} {\bibfnamefont {M.}~\bibnamefont {Viviani}}, \bibinfo {author} {\bibfnamefont {L.}~\bibnamefont {Girlanda}}, \bibinfo {author} {\bibfnamefont {A.}~\bibnamefont {Kievsky}},\ and\ \bibinfo {author} {\bibfnamefont {L.~E.}\ \bibnamefont {Marcucci}},\ }\bibfield  {title} {\bibinfo {title} {Neutron spin rotation in $\stackrel{\ensuremath{\rightarrow}}{n}\text{\ensuremath{-}}d$ scattering},\ }\href {https://doi.org/10.1103/PhysRevC.78.014002} {\bibfield  {journal} {\bibinfo  {journal} {Phys. Rev. C}\ }\textbf {\bibinfo {volume} {78}},\ \bibinfo {pages} {014002} (\bibinfo {year} {2008})}\BibitemShut {NoStop}%
\bibitem [{\citenamefont {Ramsey-Musolf}\ and\ \citenamefont {Page}(2006)}]{Ramsey-Musolf:2006vfz}%
  \BibitemOpen
  \bibfield  {author} {\bibinfo {author} {\bibfnamefont {M.~J.}\ \bibnamefont {Ramsey-Musolf}}\ and\ \bibinfo {author} {\bibfnamefont {S.~A.}\ \bibnamefont {Page}},\ }\bibfield  {title} {\bibinfo {title} {{Hadronic parity violation: A New view through the looking glass}},\ }\href {https://doi.org/10.1146/annurev.nucl.54.070103.181255} {\bibfield  {journal} {\bibinfo  {journal} {Ann. Rev. Nucl. Part. Sci.}\ }\textbf {\bibinfo {volume} {56}},\ \bibinfo {pages} {1} (\bibinfo {year} {2006})},\ \Eprint {https://arxiv.org/abs/hep-ph/0601127} {arXiv:hep-ph/0601127} \BibitemShut {NoStop}%
\bibitem [{\citenamefont {Schindler}\ and\ \citenamefont {Springer}(2013)}]{Schindler2013}%
  \BibitemOpen
  \bibfield  {author} {\bibinfo {author} {\bibfnamefont {M.~R.}\ \bibnamefont {Schindler}}\ and\ \bibinfo {author} {\bibfnamefont {R.~P.}\ \bibnamefont {Springer}},\ }\bibfield  {title} {\bibinfo {title} {The theory of parity violation in few-nucleon systems},\ }\href {https://doi.org/10.1016/j.ppnp.2013.05.002} {\bibfield  {journal} {\bibinfo  {journal} {Prog. Part. Nucl. Phys.}\ }\textbf {\bibinfo {volume} {72}},\ \bibinfo {pages} {1} (\bibinfo {year} {2013})}\BibitemShut {NoStop}%
\bibitem [{\citenamefont {Haxton}\ and\ \citenamefont {Holstein}(2013)}]{Haxton:2013aca}%
  \BibitemOpen
  \bibfield  {author} {\bibinfo {author} {\bibfnamefont {W.~C.}\ \bibnamefont {Haxton}}\ and\ \bibinfo {author} {\bibfnamefont {B.~R.}\ \bibnamefont {Holstein}},\ }\bibfield  {title} {\bibinfo {title} {{Hadronic Parity Violation}},\ }\href {https://doi.org/10.1016/j.ppnp.2013.03.009} {\bibfield  {journal} {\bibinfo  {journal} {Prog. Part. Nucl. Phys.}\ }\textbf {\bibinfo {volume} {71}},\ \bibinfo {pages} {185} (\bibinfo {year} {2013})},\ \Eprint {https://arxiv.org/abs/1303.4132} {arXiv:1303.4132 [nucl-th]} \BibitemShut {NoStop}%
\bibitem [{\citenamefont {Phillips}\ \emph {et~al.}(2015)\citenamefont {Phillips}, \citenamefont {Samart},\ and\ \citenamefont {Schat}}]{Phillips:2014kna}%
  \BibitemOpen
  \bibfield  {author} {\bibinfo {author} {\bibfnamefont {D.~R.}\ \bibnamefont {Phillips}}, \bibinfo {author} {\bibfnamefont {D.}~\bibnamefont {Samart}},\ and\ \bibinfo {author} {\bibfnamefont {C.}~\bibnamefont {Schat}},\ }\bibfield  {title} {\bibinfo {title} {{Parity-Violating Nucleon-Nucleon Force in the 1/$N_c$ Expansion}},\ }\href {https://doi.org/10.1103/PhysRevLett.114.062301} {\bibfield  {journal} {\bibinfo  {journal} {Phys. Rev. Lett.}\ }\textbf {\bibinfo {volume} {114}},\ \bibinfo {pages} {062301} (\bibinfo {year} {2015})},\ \Eprint {https://arxiv.org/abs/1410.1157} {arXiv:1410.1157 [nucl-th]} \BibitemShut {NoStop}%
\bibitem [{\citenamefont {Viviani}\ \emph {et~al.}(2014)\citenamefont {Viviani}, \citenamefont {Baroni}, \citenamefont {Girlanda}, \citenamefont {Kievsky}, \citenamefont {Marcucci},\ and\ \citenamefont {Schiavilla}}]{Viviani2014}%
  \BibitemOpen
  \bibfield  {author} {\bibinfo {author} {\bibfnamefont {M.}~\bibnamefont {Viviani}}, \bibinfo {author} {\bibfnamefont {A.}~\bibnamefont {Baroni}}, \bibinfo {author} {\bibfnamefont {L.}~\bibnamefont {Girlanda}}, \bibinfo {author} {\bibfnamefont {A.}~\bibnamefont {Kievsky}}, \bibinfo {author} {\bibfnamefont {L.~E.}\ \bibnamefont {Marcucci}},\ and\ \bibinfo {author} {\bibfnamefont {R.}~\bibnamefont {Schiavilla}},\ }\bibfield  {title} {\bibinfo {title} {Chiral effective field theory analysis of hadronic parity violation in few-nucleon systems},\ }\href {https://doi.org/10.1103/PhysRevC.89.064004} {\bibfield  {journal} {\bibinfo  {journal} {Phys. Rev. C}\ }\textbf {\bibinfo {volume} {89}},\ \bibinfo {pages} {064004} (\bibinfo {year} {2014})}\BibitemShut {NoStop}%
\bibitem [{\citenamefont {Gardner}\ \emph {et~al.}(2017)\citenamefont {Gardner}, \citenamefont {Haxton},\ and\ \citenamefont {Holstein}}]{Gardner:2017xyl}%
  \BibitemOpen
  \bibfield  {author} {\bibinfo {author} {\bibfnamefont {S.}~\bibnamefont {Gardner}}, \bibinfo {author} {\bibfnamefont {W.~C.}\ \bibnamefont {Haxton}},\ and\ \bibinfo {author} {\bibfnamefont {B.~R.}\ \bibnamefont {Holstein}},\ }\bibfield  {title} {\bibinfo {title} {{A New Paradigm for Hadronic Parity Nonconservation and its Experimental Implications}},\ }\href {https://doi.org/10.1146/annurev-nucl-041917-033231} {\bibfield  {journal} {\bibinfo  {journal} {Ann. Rev. Nucl. Part. Sci.}\ }\textbf {\bibinfo {volume} {67}},\ \bibinfo {pages} {69} (\bibinfo {year} {2017})},\ \Eprint {https://arxiv.org/abs/1704.02617} {arXiv:1704.02617 [nucl-th]} \BibitemShut {NoStop}%
\bibitem [{\citenamefont {de~Vries}\ \emph {et~al.}(2020)\citenamefont {de~Vries}, \citenamefont {Epelbaum}, \citenamefont {Girlanda}, \citenamefont {Gnech}, \citenamefont {Mereghetti},\ and\ \citenamefont {Viviani}}]{deVries:2020iea}%
  \BibitemOpen
  \bibfield  {author} {\bibinfo {author} {\bibfnamefont {J.}~\bibnamefont {de~Vries}}, \bibinfo {author} {\bibfnamefont {E.}~\bibnamefont {Epelbaum}}, \bibinfo {author} {\bibfnamefont {L.}~\bibnamefont {Girlanda}}, \bibinfo {author} {\bibfnamefont {A.}~\bibnamefont {Gnech}}, \bibinfo {author} {\bibfnamefont {E.}~\bibnamefont {Mereghetti}},\ and\ \bibinfo {author} {\bibfnamefont {M.}~\bibnamefont {Viviani}},\ }\bibfield  {title} {\bibinfo {title} {{Parity- and Time-Reversal-Violating Nuclear Forces}},\ }\href {https://doi.org/10.3389/fphy.2020.00218} {\bibfield  {journal} {\bibinfo  {journal} {Front. in Phys.}\ }\textbf {\bibinfo {volume} {8}},\ \bibinfo {pages} {218} (\bibinfo {year} {2020})},\ \Eprint {https://arxiv.org/abs/2001.09050} {arXiv:2001.09050 [nucl-th]} \BibitemShut {NoStop}%
\bibitem [{\citenamefont {Gardner}\ and\ \citenamefont {Muralidhara}(2023)}]{Gardner_prc:2022dwi}%
  \BibitemOpen
  \bibfield  {author} {\bibinfo {author} {\bibfnamefont {S.}~\bibnamefont {Gardner}}\ and\ \bibinfo {author} {\bibfnamefont {G.}~\bibnamefont {Muralidhara}},\ }\bibfield  {title} {\bibinfo {title} {{Toward a unified treatment of ${\Delta}S=0$ parity violation in low-energy nuclear processes}},\ }\href {https://doi.org/10.1103/PhysRevC.107.055501} {\bibfield  {journal} {\bibinfo  {journal} {Phys. Rev. C}\ }\textbf {\bibinfo {volume} {107}},\ \bibinfo {pages} {055501} (\bibinfo {year} {2023})},\ \Eprint {https://arxiv.org/abs/2210.03567} {arXiv:2210.03567 [nucl-th]} \BibitemShut {NoStop}%
\bibitem [{\citenamefont {Gardner}\ and\ \citenamefont {Muralidhara}(2022)}]{Gardner_plb:2022mxf}%
  \BibitemOpen
  \bibfield  {author} {\bibinfo {author} {\bibfnamefont {S.}~\bibnamefont {Gardner}}\ and\ \bibinfo {author} {\bibfnamefont {G.}~\bibnamefont {Muralidhara}},\ }\bibfield  {title} {\bibinfo {title} {{QCD analysis of ${\Delta} S=0$ hadronic parity violation}},\ }\href {https://doi.org/10.1016/j.physletb.2022.137372} {\bibfield  {journal} {\bibinfo  {journal} {Phys. Lett. B}\ }\textbf {\bibinfo {volume} {833}},\ \bibinfo {pages} {137372} (\bibinfo {year} {2022})},\ \Eprint {https://arxiv.org/abs/2203.00033} {arXiv:2203.00033 [hep-ph]} \BibitemShut {NoStop}%
\bibitem [{\citenamefont {Muralidhara}\ and\ \citenamefont {Gardner}(2024)}]{Muralidhara:2023zcs}%
  \BibitemOpen
  \bibfield  {author} {\bibinfo {author} {\bibfnamefont {G.}~\bibnamefont {Muralidhara}}\ and\ \bibinfo {author} {\bibfnamefont {S.}~\bibnamefont {Gardner}},\ }\bibfield  {title} {\bibinfo {title} {{${\Delta}S=0$ hadronic parity violation in next-to-leading order QCD: Anomalous dimension matrices and their implications}},\ }\href {https://doi.org/10.1016/j.physletb.2023.138428} {\bibfield  {journal} {\bibinfo  {journal} {Phys. Lett. B}\ }\textbf {\bibinfo {volume} {849}},\ \bibinfo {pages} {138428} (\bibinfo {year} {2024})},\ \Eprint {https://arxiv.org/abs/2311.05086} {arXiv:2311.05086 [hep-ph]} \BibitemShut {NoStop}%
\bibitem [{\citenamefont {Muralidhara}(2024)}]{Muralidhara:2024thesis}%
  \BibitemOpen
  \bibfield  {author} {\bibinfo {author} {\bibfnamefont {G.~L.}\ \bibnamefont {Muralidhara}},\ }\emph {\bibinfo {title} {{QCD Analysis of Flavor-nonchanging Hadronic Weak Processes Through Next-to-leading Order}}},\ \href {https://doi.org/10.13023/etd.2024.56} {Ph.D. thesis},\ \bibinfo  {school} {Kentucky U.} (\bibinfo {year} {2024})\BibitemShut {NoStop}%
\bibitem [{\citenamefont {Dmitriev}\ \emph {et~al.}(1983)\citenamefont {Dmitriev}, \citenamefont {Flambaum}, \citenamefont {Sushkov},\ and\ \citenamefont {Telitsin}}]{dmitriev_1983_smr17-0ng02}%
  \BibitemOpen
  \bibfield  {author} {\bibinfo {author} {\bibfnamefont {V.}~\bibnamefont {Dmitriev}}, \bibinfo {author} {\bibfnamefont {V.}~\bibnamefont {Flambaum}}, \bibinfo {author} {\bibfnamefont {O.}~\bibnamefont {Sushkov}},\ and\ \bibinfo {author} {\bibfnamefont {V.}~\bibnamefont {Telitsin}},\ }\bibfield  {title} {\bibinfo {title} {The parity violating rotation of the neutron spin in helium},\ }\href@noop {} {\bibfield  {journal} {\bibinfo  {journal} {Phys. Lett., B}\ }\textbf {\bibinfo {volume} {125}},\ \bibinfo {pages} {1} (\bibinfo {year} {1983})}\BibitemShut {NoStop}%
\bibitem [{\citenamefont {Lazauskas}\ and\ \citenamefont {Song}(2019)}]{Lazauskas:2019}%
  \BibitemOpen
  \bibfield  {author} {\bibinfo {author} {\bibfnamefont {R.}~\bibnamefont {Lazauskas}}\ and\ \bibinfo {author} {\bibfnamefont {Y.-H.}\ \bibnamefont {Song}},\ }\bibfield  {title} {\bibinfo {title} {Parity-violating neutron spin rotation in $^{4}\mathrm{He}$},\ }\href {https://doi.org/10.1103/PhysRevC.99.054002} {\bibfield  {journal} {\bibinfo  {journal} {Phys. Rev. C}\ }\textbf {\bibinfo {volume} {99}},\ \bibinfo {pages} {054002} (\bibinfo {year} {2019})}\BibitemShut {NoStop}%
\bibitem [{\citenamefont {Zaretsky}\ and\ \citenamefont {Sirotkin}(1983)}]{Zaretsky1983}%
  \BibitemOpen
  \bibfield  {author} {\bibinfo {author} {\bibfnamefont {D.~F.}\ \bibnamefont {Zaretsky}}\ and\ \bibinfo {author} {\bibfnamefont {V.~I.}\ \bibnamefont {Sirotkin}},\ }\href@noop {} {\bibfield  {journal} {\bibinfo  {journal} {Sov. J. Nucl. Phys.}\ }\textbf {\bibinfo {volume} {43}},\ \bibinfo {pages} {361} (\bibinfo {year} {1983})}\BibitemShut {NoStop}%
\bibitem [{\citenamefont {Flambaum}\ and\ \citenamefont {Sushkov}(1984)}]{Flambaum1984}%
  \BibitemOpen
  \bibfield  {author} {\bibinfo {author} {\bibfnamefont {V.}~\bibnamefont {Flambaum}}\ and\ \bibinfo {author} {\bibfnamefont {O.}~\bibnamefont {Sushkov}},\ }\bibfield  {title} {\bibinfo {title} {Parity violation in nuclear forces},\ }\href@noop {} {\bibfield  {journal} {\bibinfo  {journal} {Nuclear Physics A}\ }\textbf {\bibinfo {volume} {412}},\ \bibinfo {pages} {13} (\bibinfo {year} {1984})}\BibitemShut {NoStop}%
\bibitem [{\citenamefont {Noguera}\ and\ \citenamefont {Desplanques}(1986)}]{Noguera1986}%
  \BibitemOpen
  \bibfield  {author} {\bibinfo {author} {\bibfnamefont {S.}~\bibnamefont {Noguera}}\ and\ \bibinfo {author} {\bibfnamefont {B.}~\bibnamefont {Desplanques}},\ }\href@noop {} {\bibfield  {journal} {\bibinfo  {journal} {Nucl. Phys. A}\ }\textbf {\bibinfo {volume} {457}},\ \bibinfo {pages} {189} (\bibinfo {year} {1986})}\BibitemShut {NoStop}%
\bibitem [{\citenamefont {Flambaum}(1992)}]{Flambaum1992}%
  \BibitemOpen
  \bibfield  {author} {\bibinfo {author} {\bibfnamefont {V.~V.}\ \bibnamefont {Flambaum}},\ }\bibfield  {title} {\bibinfo {title} {Regular mechanism of parity and time invariance nonconserving effects enhancement in neutron capture and scattering near p-wave compound resonances},\ }\href@noop {} {\bibfield  {journal} {\bibinfo  {journal} {Phys. Rev. C}\ }\textbf {\bibinfo {volume} {45}},\ \bibinfo {pages} {437} (\bibinfo {year} {1992})}\BibitemShut {NoStop}%
\bibitem [{\citenamefont {Flambaum}(1993)}]{Flambaum1993DynamicalEO}%
  \BibitemOpen
  \bibfield  {author} {\bibinfo {author} {\bibfnamefont {V.~V.}\ \bibnamefont {Flambaum}},\ }\bibfield  {title} {\bibinfo {title} {Dynamical enhancement of weak interactions and quantum chaos},\ }\href {https://api.semanticscholar.org/CorpusID:120090009} {\bibfield  {journal} {\bibinfo  {journal} {Physica Scripta}\ }\textbf {\bibinfo {volume} {T46}},\ \bibinfo {pages} {198} (\bibinfo {year} {1993})}\BibitemShut {NoStop}%
\bibitem [{\citenamefont {Karl}\ and\ \citenamefont {Tadic}(1977)}]{Karl1977}%
  \BibitemOpen
  \bibfield  {author} {\bibinfo {author} {\bibfnamefont {G.}~\bibnamefont {Karl}}\ and\ \bibinfo {author} {\bibfnamefont {D.}~\bibnamefont {Tadic}},\ }\bibfield  {title} {\bibinfo {title} {Model for scattering with parity violation},\ }\href@noop {} {\bibfield  {journal} {\bibinfo  {journal} {Phys. Rev. C}\ }\textbf {\bibinfo {volume} {16}},\ \bibinfo {pages} {1726} (\bibinfo {year} {1977})}\BibitemShut {NoStop}%
\bibitem [{\citenamefont {Forte}(1978)}]{Forte1978}%
  \BibitemOpen
  \bibfield  {author} {\bibinfo {author} {\bibfnamefont {M.}~\bibnamefont {Forte}},\ }\bibfield  {title} {\bibinfo {title} {Coherent parity violation experiments with neutron beams},\ }\href@noop {} {\bibfield  {journal} {\bibinfo  {journal} {Inst. Phys. Conf. Ser.}\ }\textbf {\bibinfo {volume} {42}},\ \bibinfo {pages} {86} (\bibinfo {year} {1978})}\BibitemShut {NoStop}%
\bibitem [{\citenamefont {Flambaum}\ and\ \citenamefont {Mansour}(2022)}]{PhysRevC.105.015501}%
  \BibitemOpen
  \bibfield  {author} {\bibinfo {author} {\bibfnamefont {V.~V.}\ \bibnamefont {Flambaum}}\ and\ \bibinfo {author} {\bibfnamefont {A.~J.}\ \bibnamefont {Mansour}},\ }\bibfield  {title} {\bibinfo {title} {Parity and time-reversal invariance violation in neutron-nucleus scattering},\ }\href {https://doi.org/10.1103/PhysRevC.105.015501} {\bibfield  {journal} {\bibinfo  {journal} {Phys. Rev. C}\ }\textbf {\bibinfo {volume} {105}},\ \bibinfo {pages} {015501} (\bibinfo {year} {2022})}\BibitemShut {NoStop}%
\bibitem [{\citenamefont {Tadic}\ and\ \citenamefont {Barroso}(1978)}]{Tadic1978}%
  \BibitemOpen
  \bibfield  {author} {\bibinfo {author} {\bibfnamefont {D.}~\bibnamefont {Tadic}}\ and\ \bibinfo {author} {\bibfnamefont {A.}~\bibnamefont {Barroso}},\ }\href@noop {} {\bibfield  {journal} {\bibinfo  {journal} {Nuclear Physics A}\ }\textbf {\bibinfo {volume} {294}},\ \bibinfo {pages} {376} (\bibinfo {year} {1978})}\BibitemShut {NoStop}%
\bibitem [{\citenamefont {Su}\ \emph {et~al.}(2021)\citenamefont {Su}, \citenamefont {Wang}, \citenamefont {Jiang}, \citenamefont {Ji}, \citenamefont {Fadeev}, \citenamefont {Hu}, \citenamefont {Peng},\ and\ \citenamefont {Budker}}]{su_search_2021}%
  \BibitemOpen
  \bibfield  {author} {\bibinfo {author} {\bibfnamefont {H.}~\bibnamefont {Su}}, \bibinfo {author} {\bibfnamefont {Y.}~\bibnamefont {Wang}}, \bibinfo {author} {\bibfnamefont {M.}~\bibnamefont {Jiang}}, \bibinfo {author} {\bibfnamefont {W.}~\bibnamefont {Ji}}, \bibinfo {author} {\bibfnamefont {P.}~\bibnamefont {Fadeev}}, \bibinfo {author} {\bibfnamefont {D.}~\bibnamefont {Hu}}, \bibinfo {author} {\bibfnamefont {X.}~\bibnamefont {Peng}},\ and\ \bibinfo {author} {\bibfnamefont {D.}~\bibnamefont {Budker}},\ }\bibfield  {title} {\bibinfo {title} {Search for exotic spin-dependent interactions with a spin-based amplifier},\ }\href {https://doi.org/10.1126/sciadv.abi9535} {\bibfield  {journal} {\bibinfo  {journal} {Sci. Adv.}\ }\textbf {\bibinfo {volume} {7}},\ \bibinfo {pages} {eabi9535} (\bibinfo {year} {2021})}\BibitemShut {NoStop}%
\bibitem [{\citenamefont {Yan}\ \emph {et~al.}(2015)\citenamefont {Yan}, \citenamefont {Sun}, \citenamefont {Peng}, \citenamefont {Zhang}, \citenamefont {Fu}, \citenamefont {Guo},\ and\ \citenamefont {Liu}}]{yan_searching_2015}%
  \BibitemOpen
  \bibfield  {author} {\bibinfo {author} {\bibfnamefont {H.}~\bibnamefont {Yan}}, \bibinfo {author} {\bibfnamefont {G.~A.}\ \bibnamefont {Sun}}, \bibinfo {author} {\bibfnamefont {S.~M.}\ \bibnamefont {Peng}}, \bibinfo {author} {\bibfnamefont {Y.}~\bibnamefont {Zhang}}, \bibinfo {author} {\bibfnamefont {C.}~\bibnamefont {Fu}}, \bibinfo {author} {\bibfnamefont {H.}~\bibnamefont {Guo}},\ and\ \bibinfo {author} {\bibfnamefont {B.~Q.}\ \bibnamefont {Liu}},\ }\bibfield  {title} {\bibinfo {title} {Searching for new spin- and velocity-dependent interactions by spin relaxation of polarized $^{3}\mathrm{He}$ gas},\ }\href {https://doi.org/10.1103/PhysRevLett.115.182001} {\bibfield  {journal} {\bibinfo  {journal} {Phys. Rev. Lett.}\ }\textbf {\bibinfo {volume} {115}},\ \bibinfo {pages} {182001} (\bibinfo {year} {2015})}\BibitemShut {NoStop}%
\bibitem [{\citenamefont {Altunta\c{s}}\ \emph {et~al.}(2018)\citenamefont {Altunta\c{s}}, \citenamefont {Ammon}, \citenamefont {Cahn},\ and\ \citenamefont {DeMille}}]{Altuntas2018}%
  \BibitemOpen
  \bibfield  {author} {\bibinfo {author} {\bibfnamefont {E.}~\bibnamefont {Altunta\c{s}}}, \bibinfo {author} {\bibfnamefont {J.}~\bibnamefont {Ammon}}, \bibinfo {author} {\bibfnamefont {S.~B.}\ \bibnamefont {Cahn}},\ and\ \bibinfo {author} {\bibfnamefont {D.}~\bibnamefont {DeMille}},\ }\bibfield  {title} {\bibinfo {title} {Demonstration of a sensitive method to measure nuclear-spin-dependent parity violation},\ }\href {https://doi.org/10.1103/PhysRevLett.120.142501} {\bibfield  {journal} {\bibinfo  {journal} {Phys. Rev. Lett.}\ }\textbf {\bibinfo {volume} {120}},\ \bibinfo {pages} {142501} (\bibinfo {year} {2018})}\BibitemShut {NoStop}%
\bibitem [{\citenamefont {Gaul}\ \emph {et~al.}(2026)\citenamefont {Gaul}, \citenamefont {Cong},\ and\ \citenamefont {Budker}}]{Gaul2026}%
  \BibitemOpen
  \bibfield  {author} {\bibinfo {author} {\bibfnamefont {K.}~\bibnamefont {Gaul}}, \bibinfo {author} {\bibfnamefont {L.}~\bibnamefont {Cong}},\ and\ \bibinfo {author} {\bibfnamefont {D.}~\bibnamefont {Budker}},\ }\bibfield  {title} {\bibinfo {title} {Constraints on new vector boson mediated electron-nucleus interactions from spectroscopy of polar diatomic molecules},\ }\href {https://doi.org/10.1103/d19m-s856} {\bibfield  {journal} {\bibinfo  {journal} {Phys. Rev. Lett.}\ }\textbf {\bibinfo {volume} {136}},\ \bibinfo {pages} {181805} (\bibinfo {year} {2026})}\BibitemShut {NoStop}%
\bibitem [{\citenamefont {Dzuba}\ \emph {et~al.}(2017)\citenamefont {Dzuba}, \citenamefont {Flambaum},\ and\ \citenamefont {Stadnik}}]{Dzuba2017}%
  \BibitemOpen
  \bibfield  {author} {\bibinfo {author} {\bibfnamefont {V.~A.}\ \bibnamefont {Dzuba}}, \bibinfo {author} {\bibfnamefont {V.~V.}\ \bibnamefont {Flambaum}},\ and\ \bibinfo {author} {\bibfnamefont {Y.~V.}\ \bibnamefont {Stadnik}},\ }\bibfield  {title} {\bibinfo {title} {Probing low-mass vector bosons with parity nonconservation and nuclear anapole moment measurements in atoms and molecules},\ }\href {https://doi.org/10.1103/PhysRevLett.119.223201} {\bibfield  {journal} {\bibinfo  {journal} {Phys. Rev. Lett.}\ }\textbf {\bibinfo {volume} {119}},\ \bibinfo {pages} {223201} (\bibinfo {year} {2017})}\BibitemShut {NoStop}%
\bibitem [{\citenamefont {Clayburn}\ and\ \citenamefont {Hunter}(2023)}]{clayburn_using_2023}%
  \BibitemOpen
  \bibfield  {author} {\bibinfo {author} {\bibfnamefont {N.~B.}\ \bibnamefont {Clayburn}}\ and\ \bibinfo {author} {\bibfnamefont {L.~R.}\ \bibnamefont {Hunter}},\ }\bibfield  {title} {\bibinfo {title} {Using earth to search for long-range spin-velocity interactions},\ }\href {https://doi.org/10.1103/PhysRevD.108.L051701} {\bibfield  {journal} {\bibinfo  {journal} {Phys. Rev. D}\ }\textbf {\bibinfo {volume} {108}},\ \bibinfo {pages} {L051701} (\bibinfo {year} {2023})}\BibitemShut {NoStop}%
\bibitem [{\citenamefont {Wu}\ \emph {et~al.}(2023)\citenamefont {Wu}, \citenamefont {Zhang}, \citenamefont {Peng}, \citenamefont {Gong},\ and\ \citenamefont {Yan}}]{wu_new_2023}%
  \BibitemOpen
  \bibfield  {author} {\bibinfo {author} {\bibfnamefont {L.~Y.}\ \bibnamefont {Wu}}, \bibinfo {author} {\bibfnamefont {K.~Y.}\ \bibnamefont {Zhang}}, \bibinfo {author} {\bibfnamefont {M.}~\bibnamefont {Peng}}, \bibinfo {author} {\bibfnamefont {J.}~\bibnamefont {Gong}},\ and\ \bibinfo {author} {\bibfnamefont {H.}~\bibnamefont {Yan}},\ }\bibfield  {title} {\bibinfo {title} {New limits on exotic spin-dependent interactions at astronomical distances},\ }\href {https://doi.org/10.1103/PhysRevLett.131.091002} {\bibfield  {journal} {\bibinfo  {journal} {Phys. Rev. Lett.}\ }\textbf {\bibinfo {volume} {131}},\ \bibinfo {pages} {091002} (\bibinfo {year} {2023})}\BibitemShut {NoStop}%
\bibitem [{\citenamefont {Shu}\ \emph {et~al.}(2024)\citenamefont {Shu}, \citenamefont {Zhang}, \citenamefont {Chen}, \citenamefont {Luo}, \citenamefont {Deng}, \citenamefont {Xu}, \citenamefont {Duan}, \citenamefont {Fang}, \citenamefont {Cao}, \citenamefont {Hu},\ and\ \citenamefont {Zhou}}]{shu_constraint_2024}%
  \BibitemOpen
  \bibfield  {author} {\bibinfo {author} {\bibfnamefont {Y.-B.}\ \bibnamefont {Shu}}, \bibinfo {author} {\bibfnamefont {T.}~\bibnamefont {Zhang}}, \bibinfo {author} {\bibfnamefont {L.-L.}\ \bibnamefont {Chen}}, \bibinfo {author} {\bibfnamefont {Q.}~\bibnamefont {Luo}}, \bibinfo {author} {\bibfnamefont {X.-B.}\ \bibnamefont {Deng}}, \bibinfo {author} {\bibfnamefont {W.-J.}\ \bibnamefont {Xu}}, \bibinfo {author} {\bibfnamefont {X.-C.}\ \bibnamefont {Duan}}, \bibinfo {author} {\bibfnamefont {X.-T.}\ \bibnamefont {Fang}}, \bibinfo {author} {\bibfnamefont {L.-S.}\ \bibnamefont {Cao}}, \bibinfo {author} {\bibfnamefont {Z.-K.}\ \bibnamefont {Hu}},\ and\ \bibinfo {author} {\bibfnamefont {M.-K.}\ \bibnamefont {Zhou}},\ }\bibfield  {title} {\bibinfo {title} {Constraint on an exotic parity-odd spin- and velocity-dependent interaction with atom interferometer},\ }\href {https://doi.org/10.1103/PhysRevLett.133.213401} {\bibfield  {journal} {\bibinfo  {journal} {Phys. Rev. Lett.}\ }\textbf {\bibinfo {volume} {133}},\
  \bibinfo {pages} {213401} (\bibinfo {year} {2024})}\BibitemShut {NoStop}%
\bibitem [{\citenamefont {Adelberger}\ and\ \citenamefont {Wagner}(2013)}]{macro3}%
  \BibitemOpen
  \bibfield  {author} {\bibinfo {author} {\bibfnamefont {E.~G.}\ \bibnamefont {Adelberger}}\ and\ \bibinfo {author} {\bibfnamefont {T.~A.}\ \bibnamefont {Wagner}},\ }\bibfield  {title} {\bibinfo {title} {Improved limits on long-range parity-odd interactions of the neutron},\ }\href {https://doi.org/10.1103/PhysRevD.88.031101} {\bibfield  {journal} {\bibinfo  {journal} {Phys. Rev. D}\ }\textbf {\bibinfo {volume} {88}},\ \bibinfo {pages} {031101} (\bibinfo {year} {2013})}\BibitemShut {NoStop}%
\bibitem [{\citenamefont {Vasilakis}\ \emph {et~al.}(2009)\citenamefont {Vasilakis}, \citenamefont {Brown}, \citenamefont {Kornack},\ and\ \citenamefont {Romalis}}]{vasilakis_limits_2009}%
  \BibitemOpen
  \bibfield  {author} {\bibinfo {author} {\bibfnamefont {G.}~\bibnamefont {Vasilakis}}, \bibinfo {author} {\bibfnamefont {J.~M.}\ \bibnamefont {Brown}}, \bibinfo {author} {\bibfnamefont {T.~W.}\ \bibnamefont {Kornack}},\ and\ \bibinfo {author} {\bibfnamefont {M.~V.}\ \bibnamefont {Romalis}},\ }\bibfield  {title} {\bibinfo {title} {Limits on new long range nuclear spin-dependent forces set with a k-$^{3}$he comagnetometer},\ }\href {https://doi.org/10.1103/PhysRevLett.103.261801} {\bibfield  {journal} {\bibinfo  {journal} {Phys. Rev. Lett.}\ }\textbf {\bibinfo {volume} {103}},\ \bibinfo {pages} {261801} (\bibinfo {year} {2009})}\BibitemShut {NoStop}%
\bibitem [{\citenamefont {Wang}\ \emph {et~al.}(2023)\citenamefont {Wang}, \citenamefont {Huang}, \citenamefont {Guo}, \citenamefont {Jiang}, \citenamefont {Kang}, \citenamefont {Su}, \citenamefont {Qin}, \citenamefont {Ji}, \citenamefont {Hu}, \citenamefont {Peng},\ and\ \citenamefont {Budker}}]{wang_search_2023}%
  \BibitemOpen
  \bibfield  {author} {\bibinfo {author} {\bibfnamefont {Y.}~\bibnamefont {Wang}}, \bibinfo {author} {\bibfnamefont {Y.}~\bibnamefont {Huang}}, \bibinfo {author} {\bibfnamefont {C.}~\bibnamefont {Guo}}, \bibinfo {author} {\bibfnamefont {M.}~\bibnamefont {Jiang}}, \bibinfo {author} {\bibfnamefont {X.}~\bibnamefont {Kang}}, \bibinfo {author} {\bibfnamefont {H.}~\bibnamefont {Su}}, \bibinfo {author} {\bibfnamefont {Y.}~\bibnamefont {Qin}}, \bibinfo {author} {\bibfnamefont {W.}~\bibnamefont {Ji}}, \bibinfo {author} {\bibfnamefont {D.}~\bibnamefont {Hu}}, \bibinfo {author} {\bibfnamefont {X.}~\bibnamefont {Peng}},\ and\ \bibinfo {author} {\bibfnamefont {D.}~\bibnamefont {Budker}},\ }\bibfield  {title} {\bibinfo {title} {Search for exotic parity-violation interactions with quantum spin amplifiers},\ }\href {https://doi.org/10.1126/sciadv.ade0353} {\bibfield  {journal} {\bibinfo  {journal} {Sci. Adv.}\ }\textbf {\bibinfo {volume} {9}},\ \bibinfo {pages} {eade0353} (\bibinfo {year} {2023})}\BibitemShut {NoStop}%
\bibitem [{\citenamefont {Fichet}(2018)}]{PhysRevLett.120.131801}%
  \BibitemOpen
  \bibfield  {author} {\bibinfo {author} {\bibfnamefont {S.}~\bibnamefont {Fichet}},\ }\bibfield  {title} {\bibinfo {title} {Quantum forces from dark matter and where to find them},\ }\href {https://doi.org/10.1103/PhysRevLett.120.131801} {\bibfield  {journal} {\bibinfo  {journal} {Phys. Rev. Lett.}\ }\textbf {\bibinfo {volume} {120}},\ \bibinfo {pages} {131801} (\bibinfo {year} {2018})}\BibitemShut {NoStop}%
\bibitem [{\citenamefont {Brax}\ \emph {et~al.}(2018)\citenamefont {Brax}, \citenamefont {Fichet},\ and\ \citenamefont {Pignol}}]{PhysRevD.97.115034}%
  \BibitemOpen
  \bibfield  {author} {\bibinfo {author} {\bibfnamefont {P.}~\bibnamefont {Brax}}, \bibinfo {author} {\bibfnamefont {S.}~\bibnamefont {Fichet}},\ and\ \bibinfo {author} {\bibfnamefont {G.}~\bibnamefont {Pignol}},\ }\bibfield  {title} {\bibinfo {title} {Bounding quantum dark forces},\ }\href {https://doi.org/10.1103/PhysRevD.97.115034} {\bibfield  {journal} {\bibinfo  {journal} {Phys. Rev. D}\ }\textbf {\bibinfo {volume} {97}},\ \bibinfo {pages} {115034} (\bibinfo {year} {2018})}\BibitemShut {NoStop}%
\bibitem [{\citenamefont {Costantino}\ \emph {et~al.}(2020)\citenamefont {Costantino}, \citenamefont {Fichet},\ and\ \citenamefont {Tanedo}}]{Costantino:2019ixl}%
  \BibitemOpen
  \bibfield  {author} {\bibinfo {author} {\bibfnamefont {A.}~\bibnamefont {Costantino}}, \bibinfo {author} {\bibfnamefont {S.}~\bibnamefont {Fichet}},\ and\ \bibinfo {author} {\bibfnamefont {P.}~\bibnamefont {Tanedo}},\ }\bibfield  {title} {\bibinfo {title} {{Exotic Spin-Dependent Forces from a Hidden Sector}},\ }\href {https://doi.org/10.1007/JHEP03(2020)148} {\bibfield  {journal} {\bibinfo  {journal} {JHEP}\ }\textbf {\bibinfo {volume} {03}},\ \bibinfo {pages} {148}},\ \Eprint {https://arxiv.org/abs/1910.02972} {arXiv:1910.02972 [hep-ph]} \BibitemShut {NoStop}%
\bibitem [{\citenamefont {Feinberg}\ and\ \citenamefont {Sucher}(1968)}]{PhysRev.166.1638}%
  \BibitemOpen
  \bibfield  {author} {\bibinfo {author} {\bibfnamefont {G.}~\bibnamefont {Feinberg}}\ and\ \bibinfo {author} {\bibfnamefont {J.}~\bibnamefont {Sucher}},\ }\bibfield  {title} {\bibinfo {title} {Long-range forces from neutrino-pair exchange},\ }\href {https://doi.org/10.1103/PhysRev.166.1638} {\bibfield  {journal} {\bibinfo  {journal} {Phys. Rev.}\ }\textbf {\bibinfo {volume} {166}},\ \bibinfo {pages} {1638} (\bibinfo {year} {1968})}\BibitemShut {NoStop}%
\end{thebibliography}
\end{document}